\documentclass[preprint,11pt]{elsarticle}
\biboptions{sort&compress}

\usepackage{amsmath,amssymb,amsfonts}%
\usepackage{xcolor}%
\usepackage{algorithm}%
\usepackage{algorithmicx}%
\usepackage{algpseudocode}%
\usepackage{listings}%
\usepackage{subfig}
\usepackage{enumitem}
\setlist[enumerate,1]{label=\arabic*.}
\setlist[enumerate,2]{label=\roman*)}

\newcommand{\sty}[1]{\mbox{\boldmath $#1$}}

\newcommand{\fu}{\sty{ u}}
\newcommand{\ff}{\sty{ f}}
\newcommand{\fz}{\sty{ z}}
\newcommand{\fB}{\sty{ B}}
\newcommand{\fC}{\sty{ C}}
\newcommand{\fN}{\sty{ N}}
\newcommand{\fs}{\sty{ s}}

\newcommand{\fI}{\mathbb{\mathbf{I}}}
\newcommand{\fII}{\text{$\protect\fakebold{\mathbb{I}}$}}
\newcommand{\feps}{\mbox{\boldmath $\varepsilon $}}
\newcommand{\fsig}{\mbox{\boldmath $\sigma$}}

\newcommand{\sumE}{\overset{m}{\underset{e=1}{\sum}}}
\newcommand{\argmin}{\mathop{\mathrm{arg\,min}}}

\DeclareFontFamily{U}{mathx}{\hyphenchar\font45}
\DeclareFontShape{U}{mathx}{m}{n}{
	<5> <6> <7> <8> <9> <10>
	<10.95> <12> <14.4> <17.28> <20.74> <24.88>
	mathx10
}{}
\DeclareSymbolFont{mathx}{U}{mathx}{m}{n}
\DeclareMathSymbol{\bigtimes}{1}{mathx}{"91}
\algnewcommand\algorithmicbreak{\textbf{break}} 
\algnewcommand\Break{\algorithmicbreak{} }%
\algnewcommand\algorithmicdata{\textbf{Data:}}
\algnewcommand\Data{\item[\algorithmicdata{}]}%
\algrenewcommand\algorithmicrequire{\textbf{Input:}}
\algrenewcommand\algorithmicensure{\textbf{Output:}}

\usepackage{scalerel}
\newlength\bshft
\def\fakebold#1{\ThisStyle{\ooalign{$\SavedStyle#1$\cr%
			\kern-\bshft$\SavedStyle#1$\cr%
			\kern\bshft$\SavedStyle#1$}}}

\colorlet{kc}{red}
\colorlet{kh}{blue}

\journal{Computer Methods in Applied Mechanics and Engineering }

\begin{document}

\begin{frontmatter}

\title{Model-free data-driven inelasticity in Haigh-Westergaard space - a study how to obtain data points from measurements}

\author[rub]{Kerem Ciftci\corref{cor1}}
\ead{Kerem.Ciftci@rub.de}
\author[rub]{Klaus Hackl}
\ead{Klaus.Hackl@rub.de}
\cortext[cor1]{Corresponding author}
\address[rub]{Institute of Mechanics of Materials, Ruhr University Bochum, Universit\"atsstrasse 150, 44801 Bochum, Germany.}

\begin{abstract}
	Model-free data-driven computational mechanics, first proposed by Kirchdoerfer and Ortiz, replaces phenomenological models with numerical simulations based on sample data sets in strain-stress space. Recent literature extended the approach to inelastic problems using structured data sets, tangent space information, and transition rules. From an application perspective, the coverage of qualified data states and calculating the corresponding tangent space is crucial. In this respect, material symmetry significantly helps to reduce the amount of necessary data. This study applies the data-driven paradigm to elasto-plasticity with isotropic hardening. We formulate our approach employing Haigh–Westergaard coordinates, providing information on the underlying material yield surface. Based on this, we use a combined tension-torsion test to cover the knowledge of the yield surface and a single tensile test to calculate the corresponding tangent space. The resulting data-driven method minimizes the distance over the Haigh–Westergaard space augmented with directions in the tangent space subject to compatibility and equilibrium constraints. 
\end{abstract}

\begin{keyword}
	model-free data-driven \sep elasto-plasticity\sep isotropy, data reduction
\end{keyword}

\end{frontmatter}

\section{Introduction}\label{sec:intro}
In computational mechanics, the simulation of boundary value problems typically combines two different types of equations; conservation and constitutive laws. While conservation laws are derived from universal principles assuming an axiomatic character, constitutive laws are usually acquired by fitting the parameters of a model to given strain-stress data. Examples of such constitutive models can be found in \cite{timoshenko:1983}. However, the process of material modeling is often ill-posed and adds uncertainties to the solutions, especially in systems with high-dimensional complexity. The model-free data-driven method, introduced by Kirchdoerfer and Ortiz \cite{kirchdoerfer:2016}, bypasses the intermediate step of material modeling, incorporating experimental data directly into numerical calculations of boundary-value problems.
\smallskip\\
The method is elaborated for a variety of applications like non-linear elasticity \cite{kirchdoerfer:2016,kirchdoerfer:2017,conti:2018,nguyen:2018,galetzka:2020}, dynamics \cite{kirchdoerfer:2018}, finite strain \cite{platzer:2021} and material data identification \cite{stainier:2019}. Additional applications can be found in the area of non-local mechanics \cite{karapiperis:2019}, coupled electro-mechanical problems \cite{marenic:2022}, decoupled homogenization schemes \cite{zschocke:2022}, and model-driven coupling \cite{yang:2022}. An extension of the data-driven scheme has been made by using the tangent space to improve the learning of the underlying data structure. Ibañez et al. \cite{ibanez:2017,ibanez:2018} suggest a manifold learning approach mapping the data to a lower-dimensional space to use the locally linear embeddings. Eggersmann et al. \cite{eggersmann:2021} presented a second-order data-driven approach that uses tensor voting \cite{mordohai:2010} to obtain point-wise tangent spaces enabling the search for additional states close to the original data. For inelastic boundary value problems, Eggersmann et al. \cite{eggersmann:2019} include local histories in the data set to investigate materials with memory. Karapiperis et al. \cite{karapiperis:2021} have also suggested a variation of the scheme, considering multiscale modeling. In addition, the authors recently developed a method that incorporates the tangent space into the distance-minimizing data-driven formulation and classifies the underlying data structure into subsets according to various material behavior \cite{ciftci:2022}. The framework uses a parametrization of the material history and an optimal sampling of the mechanical system's state space. 
\\
Nevertheless, due to possible changes from elastic to plastic behavior as a function of the loading path, it is difficult to deal with data dependent on history as present in inelastic materials, provided one uses nearest neighbor clustering only. Eggersmann et al. \cite{eggersmann:2019} overcome this issue by accounting local histories in the data set, investigating three paradigms, i.e., materials with memory regarding the history of deformation, differential materials regarding histories of stress and strain, and history variables. However, it is still necessary to resort to additional models to analyze the evolution of internal variables. The proposed data-driven approach minimizes the distance between the evolving data set and a time-dependent constraint set. The result is a mixed formulation comprising classical and data-driven modeling approaches. Karapiperis et al. \cite{karapiperis:2021} have also proposed a variation of the scheme, considering multiscale modeling. The framework uses a parametrization of the material history and an optimal sampling of the mechanical system's state space. 
\smallskip\\
We recently developed a novel strategy that directly incorporates the tangent space into the distance-minimizing data-driven formulation and classifies the underlying data structure into subsets according to different material behavior \cite{ciftci:2022}. The former results in a significantly more compact system of equations and allows for interpolation in sparse data regions guaranteeing that the internal states cohere with the data set. Categorization into data subsets permits dealing with inelastic loading paths avoiding reliance on models for the evolution of history variables. In addition, we define transition rules mapping the system's internal states to the various subsets to operate on the data categories. As a result, the extended data-driven paradigm locates the closest point in the transitioning material data subset compatible with the problem's field equations and nearest to the local tangential direction.
\smallskip\\
Although the approach works sufficiently well for ideal data, the issue of data accessibility and its accompanying tangent space arises. In particular, data sets of inelastic materials could only be gathered by impractical sample testing encompassing a wide variety of loading directions. This paper addresses the data accessibility issue for isotropic elasto-plastic materials. The tangent space is characterized by the hardening of the material and the normal on the yield surface. For the former, we use data from a simple tensile test. For the latter, we project the data structure using Haigh-Westergaard coordinates to the octahedral plane in principal stress space. A combined tension-torsion test covers the underlying material yield surface information such that the associated yield function can be determined using a preferred approximation method, e.g., interpolation. The resulting data-driven scheme minimizes the distance to the tensile test data and determines the associated tangent stiffness in the Haigh-Westergaard space, subject to compatibility and equilibrium restrictions.\smallskip\\
\\
Section \ref{sec:ddriven} provides a general setting by introducing the definitions and derivation of the distance-minimizing data-driven computing method, including the enhancement of tangent space and transition rules. Section \ref{sec:mod_ddriven} introduces a coordinate transformation to the octahedral plane based on the Haigh-Westergaard coordinates. In addition, we propose an approach to determine the normal on the yield surface and the corresponding material stiffness tangent. Furthermore, we introduce transition rules to map the modeling points to the tangent space. Section \ref{sec:num_example} exhibits the performance of the proposed method using a numerical example involving elasto-plasticity with isotropic hardening. Finally, Section \ref{sec:conclusion} summarizes the results and suggests future research subjects.

\section{Tangent space enhanced data-driven paradigm}\label{sec:ddriven}
The following will summarize the classical data-driven computational mechanics method for the reader's convenience based on the definitions and formulations in \cite{kirchdoerfer:2016,eggersmann:2019}.
Let $\Omega \subset \mathbb{R}^d$ with $d \in \mathbb{N}$ be a discretized system consisting of $n\in \mathbb{N}$ nodes representing displacements $\fu \in \mathbb{R}^n$, which are subjected to applied forces $\ff \in \mathbb{R}^n$ resulting from distributed sources and Neumann boundary conditions. In addition, the system $\Omega$ comprises $m\in \mathbb{N}$ material states characterized by strain and stress fields  $\feps_e\in\mathbb{R}^{d_e}$ and $\fsig_e\in \mathbb{R}^{d_e}$, with $d_e \in \mathbb{N}$ being the dimension in Voigt notation at material point $e=1,\ldots,m$.
The system's internal state is subject to the compatibility and equilibrium conditions
\begin{align}
&\feps_e = \fB_e \fu, \quad \forall e=1,\ldots,m, \label{eq:constraint1}\\ 
&\sumE  w_e \fB^T_e \fsig_e = \ff,  \label{eq:constraint2}
\end{align}
with positive weights $w_e \in \mathbb{R}_+$, discrete gradient operator $\fB_e \in \mathbb{R}^{d_e\times n}$ and discrete divergence operator $\fB_e^T \in \mathbb{R}^{n\times d_e}$. Further, we define the set
\begin{align}
Z:=\bigtimes_{e=1}^m Z_e \quad
\text{with} \quad
Z_e:=\{(\feps_e, \fsig_e) \,\vert\, \feps_e, \fsig_e \in \mathbb{R}^{d_e}\},
\end{align}
where $Z_e \subseteq \mathbb{R}^{d_e} \times \mathbb{R}^{d_e}$ is the local phase space of material point $e$, and  $Z \subseteq \mathbb{R}^{md}\times\mathbb{R}^{md}$ is the global phase space of the finite system $\Omega$.
\\ \\ 
The distance-minimizing data-driven problem, introduced by \cite{kirchdoerfer:2016}, reads
\begin{equation}\label{eq:minmin}
\argmin_{\hat{\fz} \in \mathcal{D}} \argmin_{\fz\in \mathcal{C}}
d(\fz, \hat{\fz}) 
\end{equation}
where $\mathcal{C} \subset Z$ denotes the constraint set defined by
\begin{equation}
\mathcal{C} :=\Big\{\fz\in Z:\eqref{eq:constraint1}\; \text{and}\; \eqref{eq:constraint2}\Big\};
\end{equation}
containing all states fulfilling compatibility and equilibrium. The set $\mathcal{D} \subset Z$ consists of experimental measurements or results from small-scale simulations and is defined by
\begin{equation}\label{eq:dataset}
\mathcal{D} :=\bigtimes_{e=1}^{m}\mathcal{D}_e
\quad \text{with}  \quad
\mathcal{D}_e:= \{(\hat{\feps}_i, \hat{\fsig}_i) \in Z_e\}_{i=1}^{n_e}, 
\end{equation}
$n_e \in \mathbb{N}$ is the number of local data points associated with the integration point $e$. The distance $d:Z \times Z \to \mathbb{R}$ is defined by 
\begin{align}\label{eq:distance}
d(\fz, \hat{\fz}):=\sumE w_e d_e(\fz_e, \hat{\fz}_e ),
\end{align}
with local distance function $d_e:Z_e \times Z_e \to \mathbb{R}$ defined by
\begin{align}\label{eq:distance_e}
d_e(\fz_e, \hat{\fz}_e ):= 	\|\fz_e -  \hat{\fz}_e\|_e
\end{align}
metricized by the norm
\begin{equation}\label{eq:norm}
\|\fz_e\|_e := \frac{1}{2} E_e \|\feps_e\|^2_2 + \frac{1}{2} E_e^{-1} \|\fsig_e\|^2_2,
\end{equation}
where $E_e \in \mathbb{R}^+$ is a numerical scalar typically being of the type of elastic stiffness. \\

Thus, the data-driven method aims to find the closest point $\fz$ in the constraint set $\mathcal{C}$ to $\hat{\fz}$ in the material data set $\mathcal{D}$, or equivalently find the point in the data set that is closest to the constraint set. 
\subsection{Tangent space, structured data sets and transition rules}\label{subsec:extnd_ddriven}
To deal with inelastic materials, we extended the classical data-driven paradigm \eqref{eq:minmin} by tangent space information, structured data sets, and transition rules \cite{ciftci:2022}. 
The phase space collects a physical system's possible strain-stress states that a material can experience under certain conditions. The tangent space extension enables us to operate on the underlying structure of this phase space, which allows us to analyze the system's behavior in a neighborhood of a particular strain-stress point.
For this purpose, we recall the definition of the extended data set 
\begin{align}\label{eq:dataset_extended}
\mathcal{D}^{\text{ext}} = \bigtimes_{e=1}^m \mathcal{D}_e^{\text{ext}} \quad
\text{with} \quad \mathcal{D}_e^{\text{ext}}:=\{(\hat{\fz}_i, \fC_i) \,\vert\,  \hat{\fz}_i\in \mathcal{D}_e, \fC_i \in \mathbb{R}^{d_e \times d_e}_{\text{sym},+}\}_{i=1}^{n_e},
\end{align}
where $\fC_i$ represents the symmetric positive definite stiffness matrix at $\hat{\fz}_i = (\hat{\feps}_i, \hat{\fsig}_i)$, including possible inelastic effects.
Incorporating the tangent space directly into the data-driven computing method reads
\begin{align}\label{eq:minmin_ext}
\argmin_{(\hat{\fz}, \fC) \in \mathcal{D}^{\text{ext}}} \argmin_{\fz\in \mathcal{C}} d(\fz,\hat{\fz}),
\end{align} 
with data points $(\hat{\fz}, \fC) = \{(\hat{\fz}_e, \fC_e)\}_{e=1}^m$. Based on \cite{kirchdoerfer:2016}, we determine the optimal points iteratively using a fixed-point iteration expressed by
\begin{align}\label{eq:proj_DC}
(\hat{\fz}^{k+1}, \fC^{k+1}) = P_\mathcal{D}(P_\mathcal{C}(\hat{\fz}^k, \fC^k)),
\end{align}
where $k \in \mathbb{N}$ denotes the current iteration.
\\ \\
The first mapping $P_\mathcal{C}:\mathcal{D}^{\text{ext}} \to \mathcal{C}$ projects a data state $(\hat{\fz}^k, \fC^k)\in \mathcal{D}^{\text{ext}}$ to the closest point in the constraint set $\fz^k \in \mathcal{C}$. For fixed data points $\{(\hat{\fz}_e, \fC_e)\}_{e=1}^m$, e.g., from a previous iteration, the projection is performed by solving the linear equation system \cite{ciftci:2022}
\begin{equation}\label{eq:lin_eq_sys}
\left(\sumE  w_e \fB^T_e \fC_e \fB_e \right) \fu = \ff - \sumE  w_e \fB^T_e (\hat{\fsig}_e  - \fC_e \hat{\feps}_e ),
\end{equation}
and computing the corresponding strain and stress values by 
\begin{align}
\feps_e &= \fB_e \fu  && \forall e=1,\ldots,m, \label{eq:eps_e}\\
\fsig_e &=  \hat{\fsig}_e + \fC_e (\feps_e - \hat{\feps}_e) && \forall e=1,\ldots,m. \label{eq:sig_e}
\end{align}
The second projection $P_\mathcal{D}: \mathcal{C} \to \mathcal{D}^{\text{ext}}$ finds the closest state in the data set to the previously calculated state in the constraint set. We associate different tangent spaces to data points with different histories. The local material data sets $\mathcal{D}_{e}^{\text{ext}}$ are classified into subsets corresponding to different material behavior, e.g., elastic and inelastic:
\begin{align}\label{eq:data_subsets}
\mathcal{D}_e^\text{ext} =  \dot{\bigcup\limits_p}\, \mathcal{D}_e^{\text{ext},\,p} \quad \text{with } p = \{\text{elastic},\,\text{inelastic}\}. 
\end{align}
Based on the classification, transition rules map the modeling points to the various subsets. Thus, the closest point projection $P_\mathcal{D}$ is done by minimizing the local distances $d_e$ for the material states $\fz_e^k$ subject to data subset $\mathcal{D}_e^{\text{ext},\,p}$. In other words, a nearest-neighbor problem has to be solved to find the data point $(\hat{\fz}_e^{k+1}, \fC_e^{k+1}) \in \mathcal{D}_e^{\text{ext},\,p}$ closest to $\fz_e^{k+1}$ regarding the metric \eqref{eq:distance}. 
\\
In \cite{ciftci:2022}, we derived such a projection for the case of elasto-plasticity with isotropic hardening. A yield condition governs the kinetics of elasto-plasticity by
\begin{equation}\label{eq:stress_comp}
\sigma_\mathrm{com}(\fsig) \leq \sigma_\mathrm{y},
\end{equation}
where $\sigma_\mathrm{com}(\fsig)$ is comparison stress dependent on the current stress state and
$ \sigma_\mathrm{y}$ denotes the yield stress, a material property depending on the loading history in the case of isotropic hardening.
For fixed modeling points $\{\fz_e\}_{e=1}^m$ e.g. achieved from the linear equation system~\eqref{eq:lin_eq_sys}, the mapping $P_\mathcal{D}$ can be performed for material state $e=1,\ldots,m$ by:
\begin{enumerate}
\item check yield condition and assign index 
\begin{align}
	p =
	\begin{cases}
		\text{elastic}, &\text{if }
		\sigma_\mathrm{com} ({\fsig_e}) <\alpha_{\mathrm{y},e} \\ 
		\text{inelastic}, &\text{otherwise};
	\end{cases}
\end{align}
\item\label{itm:yield} if $p \equiv \text{inelastic}$, set new yield stress 
\begin{align}
	\sigma_{\mathrm{y},e}  \equiv \sigma_\mathrm{com} ({\fsig_e});
\end{align}
\item \label{itm:nneighbor} find closest data point $(\hat{\fz}_e, \fC_e)$ to modeling point $\fz_e$ by
\begin{align}
	\min\{ d_e(\fz_e,\hat{\fz}_{e})\,\vert\, (\hat{\fz}_e, \fC_e)\in \mathcal{D}_e^{\text{ext},\, p} \}.
\end{align}
\end{enumerate}
The first step maps the modeling points to the corresponding data sets; steps \ref{itm:yield} and \ref{itm:nneighbor} define a new yield limit and find the closest data point inside these sets for the next loading increment. \\ \\
Studying the data-driven approach for ideal data, we realize that the issue of the accessibility of data and its corresponding tangent space is crucial. While $100$ data points are sufficient in the $1D$ case, $100^3$ are needed for the $2D$ and $100^6$ for the $3D$ case \cite{eggersmann:2021}. Considering the latter case, the local data sets consist of strain and stress pairs $(\hat{\feps}_i, \hat{\fsig}_i)$ with $12$ independent components. By considering symmetry, the corresponding tangents $\fC_i$ have $21$ independent components. This fact raises the question of obtaining suitable data sets from measurements to cover the phase space and determine the consistent tangent space. Especially data sets for the simulation of inelastic material behavior could only be obtained by impracticable sample tests covering a variety of loading paths. To overcome this issue, we will introduce the octahedral plane in Haigh-Westergaard coordinates providing a transformation of the mapping $P_\mathcal{D}$ in principal stress space. We will show that data sets obtained from combined tension-torsion and single tension tests are sufficient for the data-driven modeling of isotropic elasto-plasticity. 

\section{Data-driven paradigm in octahedral plane}\label{sec:mod_ddriven} 
In this section, we suggest a way to obtain data from measurements insofar as we have to modify the data-driven fixed-point method's projection $P_\mathcal{D}$.

\subsection{Preliminaries for tangents of isotropic elasto-plastic bodies}
We start by introducing the constitutive relation for isotropic elasto-plastic materials
\begin{align}\label{eq:const_ep}
\fsig = \lambda\,\mathrm{tr}(\feps)\fI + 2\mu (\feps - \feps^p),
\end{align}
with strain $\feps$, stress $\fsig$, plastic strain $\feps^p$, Lamé constants $\lambda, \mu$ and second-rank identity tensor $\fI$. We employ strain and stress in means of tensors instead of Voigt form, i.e., $(\feps, \fsig) \in \mathbb{R}^{d \times d} \times \mathbb{R}^{d \times d}$.  
Taking the derivative of Eq.~\eqref{eq:const_ep} and making use of plasticity theory, it follows
\begin{align}\label{eq:const_der}
\dot{\fsig}  &=  \lambda\,\mathrm{tr}(\dot{\feps})\fI + 2\mu (\dot{\feps} - \dot{\feps}^p) = \fC:\dot{\feps},
\end{align}
with  tangent operator
\begin{align}\label{eq:tangent_eq}
\fC= \lambda \fI \otimes \fI + 2\mu \fII - \gamma  \fN \otimes \fN.
\end{align}	
In this context, $\fII$ is the symmetric part of the fourth-rank identity tensor, $\gamma \in \mathbb{R}_{\geq 0}$ is a parameter depending on the hardening and $\fN$ is the normal given by
\begin{align}\label{eq:normal}
\fN =  \Big\| \frac{\partial \Phi}{\partial \fsig} \Big\|^{-1} \frac{\partial \Phi}{\partial \fsig}, 
\end{align}
with $\Phi$ being a continuously differentiable yield function. The Lamé constants can be calculated using a simple tension test. Alternatively we can decompose the tangent given in Eq.~\eqref{eq:tangent_eq} by $\fC= \fC^\text{el} - \fC^\text{pl}$ with
\begin{align}
\fC^\text{el}:= \lambda \fI \otimes \fI + 2\mu \fII \quad \text{and} \quad  \fC^\text{pl}:= \gamma \fN \otimes \fN.
\end{align}
The matrix $\fC^\text{el}$ can be calculated using a principal component analysis applied to the elastic part of data \cite{eggersmann:2021_2}. Thus, the main task to obtain tangent $\fC$ at a fixed point  $(\feps, \fsig)$ is the computation of $\fC^\text{pl}$ depending only on normal $\fN$ and parameter $\gamma \in \mathbb{R}_{\geq 0}$. 

\subsection{Normal vector in the octahedral plane based on Haigh–Westergaard coordinates}\label{subsec:normal_hw_coord}
\begin{figure}[H]
\centering
\includegraphics[scale=1.0]{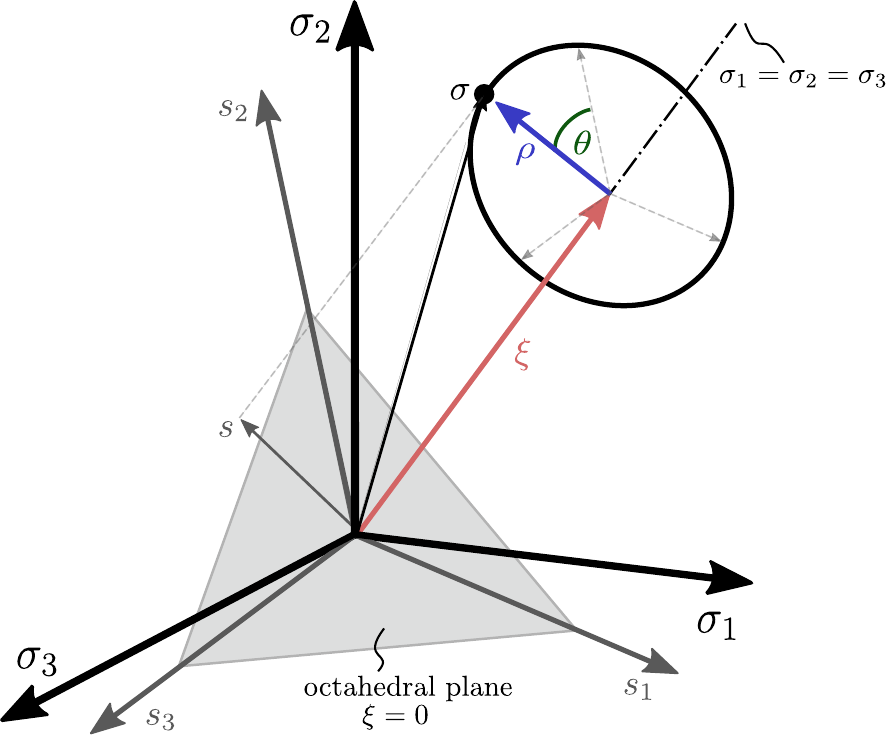}
\caption{Visualisation of a stress tensor $\fsig$ and its
	deviatoric part in the Haigh-Westergaard stress space.}
\label{fig:hw_coord}
\end{figure}

\begin{figure}[h]
\centering
\includegraphics[scale=0.6]{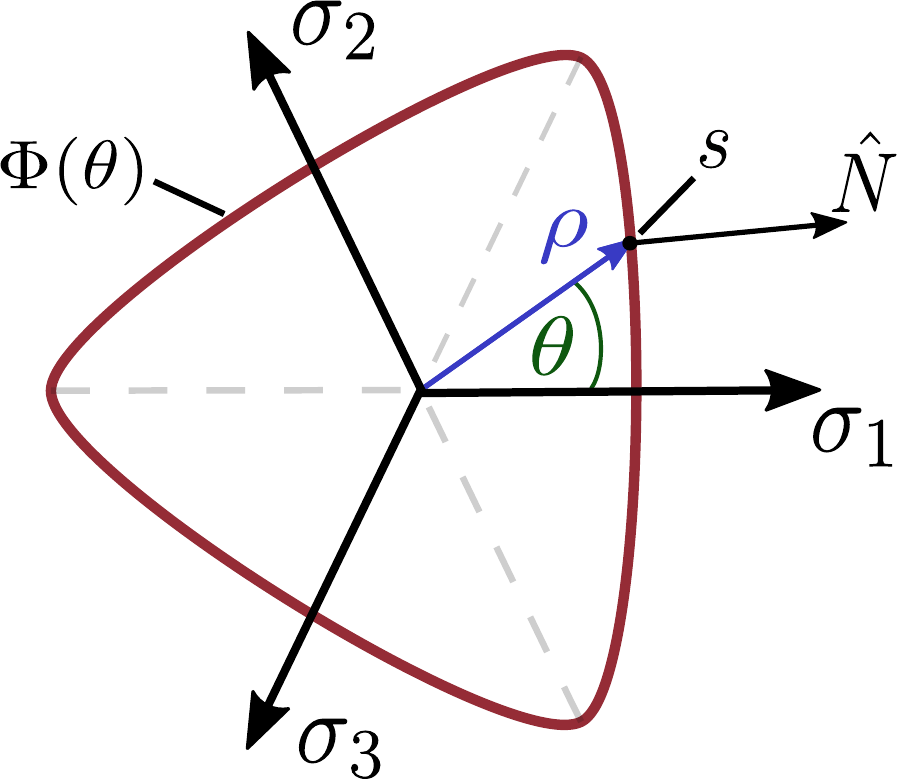}
\caption{Schematic illustration of a normal vector $\hat{\fN}$ at a random point $\fs$ of a parametrized yield surface $\Phi$ in Haigh-Westergaard coordinates}
\label{fig:yield_normal}
\end{figure}

The normal $\fN$ on the yield surface $\Phi$ at a fixed point is orthogonal to the corresponding tangent vector at this point, which can be obtained by differentiating the corresponding position vector on the surface. Instead of performing the calculations in the Cartesian coordinate system, we transform the position vector to a curvilinear system, i.e., principal stress space $(\sigma_1,\sigma_2,\sigma_3)$, where each component of the vector is expressed by cylindrical coordinates determining the surface. For this purpose, we introduce the Haigh–Westergaard coordinates $(\xi,\rho,\theta )$ describing a cylindrical coordinate system within principal stress space. Coordinate $\xi$ is the projection on the vector $(1,1,1)$ of the hydrostatic axis, and $(\rho,\theta)$ are polar coordinates in the deviatoric plane that is orthogonal to the hydrostatic axis \cite{mentrey:1995,jiang:2020}. The coordinates can be computed using invariants of the stress tensor $\fsig$ and its deviator $\fs$ defined as
\begin{align}\label{eq:invariants}
J_1 &= \mathrm{tr}(\fsig), \\
J_2 &= \frac{1}{2}\left[\mathrm{tr}(\fsig^2) - \frac{1}{3}\mathrm{tr}(\fsig)^2\right] =  \frac{1}{2}\mathrm{tr}(\fs \cdot \fs), \\
J_3 &= \det(\fs).
\end{align}
Based on this, the Haigh–Westergaard coordinates $(\xi ,\rho,\theta )$ can be obtained by
\begin{align}
\xi &= \frac{J_1}{\sqrt{3}} = \frac{\sigma_1+\sigma_2+\sigma_3}{\sqrt{3}}, \label{eq:hw_coord1}\\
\rho &= \sqrt{2J_2} = \frac{1}{3}\sqrt{(\sigma_1-\sigma_2)^2+(\sigma_2-\sigma_3)^2+(\sigma_3-\sigma_1)^2}, \\
\theta &= \frac{1}{3}\arccos\left(\frac{3\sqrt{3}}{2} J_3 J_2^{-3/2}\right) \nonumber \\ &= \arccos\left(\frac{2\sigma_1-\sigma_2-\sigma_3}{\sqrt{2}\sqrt{(\sigma_1-\sigma_2)^2+(\sigma_2-\sigma_3)^2+(\sigma_3-\sigma_1)^2}}\right), \label{eq:hw_coord2}
\end{align}
with $\theta \in [0, \frac{\pi}{3}]$. A point $\fsig$ can then be expressed in terms of the coordinates $(\xi,\rho,\theta)$ as
\begin{align}
\fsig= \begin{pmatrix}
	\sigma_1 \\ \sigma_2 \\ \sigma_3 
\end{pmatrix} = 	\frac{\xi}{\sqrt{3}} \begin{pmatrix}
	1 \\ 1 \\ 1
\end{pmatrix} +  \sqrt{\frac{2}{3}}\rho \begin{pmatrix}
	\cos(\theta)\\
	\cos\left(\theta-\frac{2\pi}{3}\right)\\
	\cos\left(\theta+\frac{2\pi}{3}\right)
\end{pmatrix},
\end{align}
with principal stresses $\sigma_1 \geq \sigma_2 \geq \sigma_3$. For $\xi \equiv 0$, the resulting plane, known as the deviatoric or octahedral plane, is a subspace of Haigh–Westergaard space given by $(\rho, \theta)$. An illustration of the coordinates and the resulting plane is given in Fig. \ref{fig:hw_coord}.
In addition, the intersection of the yield surface with the octahedral plane is given by
\begin{align}\label{eq:rho_alpha}
\rho(\theta) = \alpha \Phi(\theta),
\end{align}
with $\alpha \in \mathbb{R}_{\geq 1}$ describing isotropic hardening. In particular, for $\alpha =1$, the equation \eqref{eq:rho_alpha} represents the initial yield surface. As mentioned before, the normal to this surface at a fixed point $\fsig$ is perpendicular to any tangent vector at this point (see Fig. \ref{fig:yield_normal}).
Regarding the Haigh-Westergaard space, the tangent space is a $2$-dimensional plane whose basis consists of two tangent vectors given by
\begin{align}\label{eq:tangent_vec}
\bold t_1 = \begin{pmatrix}
	1 \\ 1 \\ 1
\end{pmatrix} 
\quad \text{and} \quad 
\bold t_2 = \frac{\partial\fs}{\partial\theta},
\end{align}
where $\fs$ is the deviatoric part of point $\fsig$ defined in Haigh-Westergaard coordinates as
\begin{align}\label{eq:dev_hw}
\fs = \sqrt{\frac{2}{3}}\rho(\theta)
\begin{pmatrix}
	\cos\left(\theta\right) \\ 
	\cos\left(\theta - \frac{2\pi}{3}\right) \\
	\cos\left(\theta + \frac{2\pi}{3}\right)
\end{pmatrix}.
\end{align}
The derivative of $\fs$ with respect to $\theta$ is given by 
\begin{align}\label{eq:dev_der_hw}
\frac{\partial\fs}{\partial\theta} 
= \sqrt{\frac{2}{3}}
\rho(\theta)\begin{pmatrix}
	-\sin\left(\theta\right)\\ 
	\cos\left(\frac{\pi}{6}-\theta\right)\\ 
	-\cos\left(\frac{\pi}{6}+\theta\right)
\end{pmatrix}
+  \sqrt{\frac{2}{3}}
\rho^\prime(\theta)	\begin{pmatrix}
	\cos\left(\theta\right) \\ 
	\cos\left(\theta - \frac{2\pi}{3}\right) \\
	\cos\left(\theta + \frac{2\pi}{3}\right)
\end{pmatrix}.
\end{align}
Based on Szeptyński \cite{szeptynski:2014}, the normal vector can then be calculated with the tangential vectors given in Eq.~\eqref{eq:tangent_vec} by
\begin{align}\label{eq:normal_t1t2}
\hat{\fN} = -\frac{\bold t_1 \times \bold t_2}{\|\bold t_1 \times \bold t_2\|_2},
\end{align}
where $\|\cdot\|_2$ denotes the well-known Euclidean norm. Substitution of Eq.~\eqref{eq:dev_der_hw} into Eq.~\eqref{eq:tangent_vec} and using trigonometric identities yields for the cross product of the tangent vectors
\begin{align}\label{eq:cp_calc}
\bold t_1 \times \bold t_2 =& 
\sqrt{\frac{2}{3}} 
\rho(\theta)
\begin{pmatrix}
	\cos\left(\frac{\pi}{6} - \theta\right) + 	\cos\left(\frac{\pi}{6} + \theta\right) \\
	\sin\left(\theta\right) - \cos\left(\frac{\pi}{6} + \theta\right) \\
	-\sin\left(\theta\right) -\cos\left(\frac{\pi}{6} - \theta\right) 
\end{pmatrix}
\\
&+\sqrt{\frac{2}{3}} 
\rho^\prime(\theta)
\begin{pmatrix}
	\sin\left(\frac{\pi}{6} + \theta\right) - \sin\left(\frac{\pi}{6} - \theta\right) \\
	-\cos\left(\theta\right) - \sin\left(\frac{\pi}{6} + \theta\right) \\
	\cos\left(\theta\right) + \sin\left(\frac{\pi}{6} - \theta\right)
\end{pmatrix}
\notag\\
=& \sqrt{2}
\rho(\theta)
\begin{pmatrix}
	\cos\left(\theta\right) \\
	-\sin\left(\frac{\pi}{6} - \theta\right) \\
	-\sin\left(\frac{\pi}{6} + \theta\right)
\end{pmatrix}
+ \sqrt{2}
\rho^\prime(\theta)
\begin{pmatrix}
	\sin\left(\theta\right) \\
	-\cos\left(\frac{\pi}{6}-\theta\right)  \\
	\cos\left(\frac{\pi}{6}+\theta\right)
\end{pmatrix}\\
=& 
\sqrt{3}\fs+ \sqrt{2}\rho^\prime(\theta)
\begin{pmatrix}
	\sin\left(\theta\right) \\
	-\cos\left(\frac{\pi}{6}-\theta\right)  \\
	\cos\left(\frac{\pi}{6}+\theta\right)
\end{pmatrix}\label{eq:cp}.
\end{align}
Applying the definition of Euclidean norm to Eq.~\eqref{eq:cp} and making use of trigonometric identities, it follows
\begin{align}\label{eq:cp_norm_calc}
\|\bold t_1 \times \bold t_2 \|_2^2 =& 
\left\lVert\sqrt{3}\fs+ \sqrt{2}\rho^\prime(\theta)
\begin{pmatrix}
	\sin\left(\theta\right) \\
	-\cos\left(\frac{\pi}{6}-\theta\right)  \\
	\cos\left(\frac{\pi}{6}+\theta\right)
\end{pmatrix}\right\rVert_2^2 \\
=&
\left(\sqrt{2}\rho(\theta)\cos\left(\theta\right) + \sqrt{2}\rho^\prime(\theta)\sin\left(\theta\right)\right)^2 \\
&+ \left(-\sqrt{2}\rho(\theta)\sin\left(\frac{\pi}{6} - \theta\right) -\sqrt{2}\rho^\prime(\theta)\cos\left(\frac{\pi}{6}-\theta\right)\right)^2 \notag
\notag\\
&+ \left(-\sqrt{2}\rho(\theta)\sin\left(\frac{\pi}{6} - \theta\right) -\sqrt{2}\rho^\prime(\theta)\cos\left(\frac{\pi}{6}-\theta\right)\right)^2 
\notag\\
=&
2\rho(\theta)^2\left(\cos\left(\theta\right)^2 + \sin\left(\frac{\pi}{6} - \theta\right)^2 + \sin\left(\frac{\pi}{6} + \theta\right)^2\right)  
\\
&+ 2\rho^\prime(\theta)^2\left(\sin\left(\theta\right)^2 + \cos\left(\frac{\pi}{6}-\theta\right)^2 +\cos\left(\frac{\pi}{6}+\theta\right)^2  \right) 
\notag\\
&+ 2\rho(\theta)\rho^\prime(\theta)\left(2\cos\left(\theta\right)\sin\left(\theta\right) + \cos\left( \frac{\pi}{6} + 2\theta\right)  - \cos\left( \frac{\pi}{6} - 2\theta\right)\right)
\notag\\
=&
2\rho(\theta)^2\left(\cos\left(\theta\right)^2 + \frac{3}{2}\sin\left(\theta\right)^2 + \frac{1}{2}\cos\left(\theta\right)^2\right) 
\\
&+2\rho^\prime(\theta)^2\left(\sin\left(\theta\right)^2 + \frac{1}{2}\sin\left(\theta\right)^2 + \frac{3}{2}\cos\left(\theta\right)^2 \right) 
\notag\\
&+2\rho(\theta)\rho^\prime(\theta)\left(2\cos\left(\theta\right)\sin\left(\theta\right)-2\cos\left(\theta\right)\sin\left(\theta\right)\right)
\notag\\
&=3\left(\rho(\theta)^2 + \rho^\prime(\theta)^2\right)\label{eq:cp_norm}.
\end{align}
The normal vector in principal stress space can then be expressed in Haigh-Westergaard coordinates by substituting Eq.~\eqref{eq:cp} and Eq.~\eqref{eq:cp_norm} into Eq.~\eqref{eq:normal_t1t2} yielding	
\begin{align}\label{eq:normal_hw}
\hat{\fN}(\theta) = \frac{1}{\sqrt{\rho(\theta)^2 + \rho^\prime(\theta)^2}} \left[ \fs + \sqrt{\frac{2}{3}} \rho^\prime(\theta) \begin{pmatrix}
	\sin\left(\theta\right) \\ -\cos\left(\frac{\pi}{6}-\theta\right) \\ \cos\left(\frac{\pi}{6}+\theta\right)
\end{pmatrix}\right].
\end{align}
The normal $\hat{\fN}$ can be transformed back to Cartesian coordinate system by $\fN= \bold T \cdot \mathrm{diag}(\hat{\fN}) \cdot \bold T^{-1}$, where $\bold T \in \mathbb{R}^{3\times 3}$  consists of eigenvectors corresponding to the eigenvalues $\sigma_{1} \geq \sigma_{2}\geq\sigma_{3}$ of $\fsig$ and satisfies the transformation to principal stress space $\fsig = \bold T \cdot \mathrm{diag}(\sigma_{1},\sigma_{2},\sigma_{3})\cdot \bold T^{-1}$. 

\subsection{Data enforced tangent}\label{subsec:data_enforced_tangent} 
\begin{figure}[h]
\centering
\includegraphics[scale=0.6]{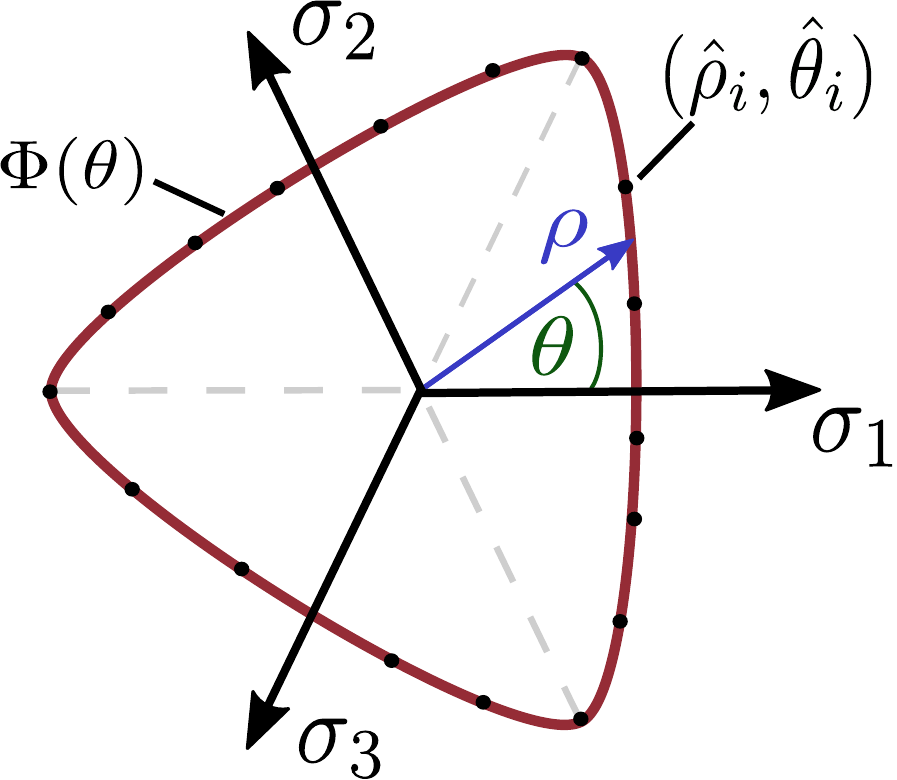}
\caption{Schematic illustration of data points $(\hat{\rho}_i, \hat{\theta}_i)$ which are gained by combined tensile-torsion test lying on the initial yield surface $\Phi(\theta)$.}
\label{fig:yield_points}
\end{figure}

Recalling the definition of the tangent 
\begin{align}
\fC= \lambda \fI \otimes \fI + 2\mu \fII - \gamma  \fN \otimes \fN,
\end{align}	
we found a formula for the normal $\fN$ in principal stress space by Eq.~\eqref{eq:normal_hw} provided we have information about the initial yield surface $\Phi(\theta)$, cf. Eq.~\eqref{eq:rho_alpha}. To address the latter, we assume a combined tensile-torsion test resulting in stress fields of the form
\begin{align}\label{eq:data_tensile_torsion}
\hat{\fsig}_i = 
\begin{pmatrix}
	\hat{\sigma}_{11, i} & 0 & 0 \\
	0 & 0 & \hat{\sigma}_{23, i} \\
	0 & \hat{\sigma}_{23, i} & 0
\end{pmatrix}.
\end{align}
By variation of the ratio $\hat{\sigma}_{23}/\hat{\sigma}_{11}$, it is possible to obtain data points \\ $\{(\rho_i, \theta_i)\}_{i=1}^{n_e}$ lying on the initial yield surface, illustrated in Fig. \ref{fig:yield_points}. By this means, it is possible to approximate the yield function $\Phi(\theta)$ by choosing an appropriate interpolation method, e.g., multilinear polynomial, spline, or nearest-neighbor interpolation.
\\
Thus, the remaining task to calculate the complete tangent operator $\fC$ is to determine parameter $\gamma \in\mathbb{R}_{\geq 0}$ depending on the hardening variable $\alpha$, i.e., $\gamma \equiv \gamma(\alpha)$. Regarding this, we assume a data set  $\{(\hat{\feps}_i, \hat{\fsig}_i)\}_{i=1}^{n_e}$  consisting of strain stress pairs of the form
\begin{align}\label{eq:data_tensile}
(\hat{\feps}_i, \hat{\fsig}_i) = \left(
\begin{pmatrix}
	\hat{\varepsilon}_{11, i} & 0 & 0\\ 
	0 & \hat{\varepsilon}_{22, i} & 0 \\ 
	0 & 0 & \hat{\varepsilon}_{22, i}
\end{pmatrix}, 
\begin{pmatrix}
	\hat{\sigma}_{11, i} & 0 & 0\\
	0 & 0 & 0 \\
	0 & 0 & 0
\end{pmatrix}\right),
\end{align}
which can be achieved by a simple tensile test. In addition, we calculate strain and stress increments $(\Delta \hat{\feps}_i, \Delta \hat{\fsig}_i) =	(\hat{\feps}_i, \hat{\fsig}_i) - (\hat{\feps}_{i-1}, \hat{\fsig}_{i-1})$ having the form
\begin{align}
(\Delta \hat{\feps}_i, \Delta \hat{\fsig}_i) =
\left(
\begin{pmatrix}
	\Delta\hat{\varepsilon}_{11, i} & 0 & 0\\ 
	0 & \Delta\hat{\varepsilon}_{22, i} & 0 \\ 
	0 & 0 & \Delta\hat{\varepsilon}_{22, i}
\end{pmatrix}, 
\begin{pmatrix}
	\Delta\hat{\sigma}_{11, i} & 0 & 0\\
	0 & 0 & 0 \\
	0 & 0 & 0
\end{pmatrix}\right).
\end{align}
Substitution of the increments into the derivative relation \eqref{eq:const_der} results in
\begin{align}\label{eq:const_incr}
\Delta \hat{\fsig}_i = \lambda \mathrm{tr}(	\Delta\hat{\feps}_i)\fI + 2\mu \Delta\hat{\feps}_i - \gamma_i \fN(\theta_i) \otimes \fN(\theta_i),
\end{align}
representing an implicit equation for parameter $\gamma_i$ for $i=1,\ldots,n_e$. As a consequence, calculating the normals $\fN(\theta_i)$ by Eq.~\eqref{eq:normal} and parameters $\gamma_i$ by Eq.~\eqref{eq:const_incr}, we can determine the tangents $\fC_i$ for the strain stress data in \eqref{eq:data_tensile}. In the following, we will exploit how to use this to calculate the tangent for a general given modeling point arising in the context of the data-driven approach. 

\subsection{Adapted data-driven projection $P_\mathcal{D}$}\label{subsec:datadriven_mod}
As mentioned in Section \ref{subsec:extnd_ddriven}, the issue of accessibility of data and its corresponding tangent space is crucial. In Section \ref{subsec:normal_hw_coord} and \ref{subsec:data_enforced_tangent}, we introduced a way to determine the tangent $\fC$ for strain stress data obtained by a simple tensile test \eqref{eq:data_tensile} using the Haigh-Westergaard space. To exploit this approach for the data-driven method, we recall the fixed-point mapping $P_\mathcal{D}P_\mathcal{C}$ in Eq.~\eqref{eq:proj_DC}, where $P_\mathcal{C}$ projects a data state $(\hat{\fz}, \fC)\in \mathcal{D}^{\text{ext}}$ to the closest point in the constraint set $\fz \in \mathcal{C}$ and computes the corresponding strain and stress values by Eq.~\eqref{eq:eps_e} and Eq.~\eqref{eq:sig_e}. Additionally, projection $P_\mathcal{D}$ finds the closest state $(\hat{\fz}, \fC)$ in data set $ \mathcal{D}^{\text{ext}}$ to the calculated material state $\fz$. While $P_\mathcal{C}$ only consists of solving linear equation system \eqref{eq:lin_eq_sys} and therefore cannot be modified in general, projection $P_\mathcal{D}$ is used to adapt the nearest neighbor search in principal stress space. \\ \\
We start by assigning the data set \eqref{eq:dataset} by
\begin{align}\label{dataset_mod}
\mathcal{D} = \bigtimes_{e=1}^m \mathcal{D}_e \quad
\text{with} \quad \mathcal{D}_e:=\{(\hat{\feps}_i, \hat{\fsig}_i)\}_{i=1}^{n_e},
\end{align}
with strain and stress points obtained by the tensile test described in \eqref{eq:data_tensile}. Based on this, we calculate the invariants 
\begin{align}
J_{2,i} &= \frac{\hat{\sigma}_{11, i}^2}{3},\\
J_{3,i} &= \frac{2\hat{\sigma}_{11, i}^3}{27},
\end{align}
providing the corresponding Haigh-Westergaard coordinates \eqref{eq:hw_coord1}-\eqref{eq:hw_coord2} by
\begin{align}
\hat{\rho}_i &= \sqrt{\frac{2}{3}} \|	\hat{\sigma}_{11, i}\|, \\
\cos\left(3\hat{\theta}_i\right) &= \frac{3\sqrt{3}}{2}\cdot\frac{2\hat{\sigma}_{11, i}^3}{27}\cdot 3^{3/2} \hat{\sigma}_{11, i}^{-3} = 1. \label{eq:cos_tmp}
\end{align}
Using that Equation~\eqref{eq:cos_tmp} yields $\hat{\theta}_i = 0$ and substituting this into relation \eqref{eq:rho_alpha}, we have an equation for the hardening variable given by
\begin{align}
\hat{\alpha}_i = \frac{\hat{\rho}_i}{\Phi(0)} = \sqrt{\frac{2}{3}} \frac{\|\hat{\sigma}_{11, i}\|}{\Phi(0)},
\end{align}
for $i=1,\ldots,n_e$. Accordingly, we redefine the extended data set \eqref{eq:dataset_extended} by 
\begin{align}\label{eq:dataset_extended_mod}
\mathcal{D}^{\text{ext}} = \bigtimes_{e=1}^m \mathcal{D}_e^{\text{ext}} \quad
\text{with} \quad \mathcal{D}_e^{\text{ext}}:=\{(\Delta \hat{\feps}_i, \Delta \hat{\fsig}_i),	\hat{\alpha}_i\}_{i=1}^{n_e},
\end{align}
with strain and stress increments $(\Delta \hat{\feps}_i, \Delta \hat{\fsig}_i) = 	(\hat{\feps}_i, \hat{\fsig}_i) - (\hat{\feps}_{i-1}, \hat{\fsig}_{i-1})$. The hardening variable $\alpha_i$ can replace the comparison stress in the transition rules of the data-driven mapping $P_\mathcal{D}$. Following the procedure of \cite{ciftci:2022}, the data sets $\mathcal{D}_{e}^{\text{ext}}$ are classified into subsets corresponding to elastic and inelastic material behavior according to Eq.~\eqref{eq:data_subsets}.
%\begin{align}\label{eq:data_subsets_mod}
%	\mathcal{D}_e^\text{ext} =  \dot{\bigcup\limits_p}\, \mathcal{D}_e^{\text{ext},\,p} \quad \text{with } p = \{\text{elastic},\,\text{inelastic}\}. 
%\end{align}
In the context of the kinetics of elasto-plasticity, the comparison stress is defined as
\begin{align}\label{eq:stress_comp_mod}
\sigma_{\mathrm{com}}(\fsig) = \frac{\rho}{\Phi(\theta)} = \alpha,
\end{align}
where the Haigh-Westergaard coordinates $\rho, \theta$ depend on $\fsig$ and $\Phi(\theta)$ is a given approximation of the yield surface, cf. Section \ref{subsec:data_enforced_tangent}. Thus, the yield condition reads
\begin{align}
\sigma_{\mathrm{com}}(\fsig) \leq \alpha_y,
\end{align}
with $\alpha_y \in \mathbb{R}_{\geq 1}$ denoting the hardening parameter. 
Hence, for fixed modeling points $\{\fz_e\}_{e=1}^m$ the mapping $P_\mathcal{D}:\mathcal{C}\to \mathcal{D}^{\text{ext}}$ can be adapted for material state $e=1,\ldots,m$ by:
\\
\hrule
\begin{enumerate}
\item\label{itm:hw_coord} determine the Haigh-Westergaard coordinates $\rho_e,\theta_e$
\item\label{itm:yield_cond} check yield condition and assign index 
\begin{align}
	p =\begin{cases}
		\text{elastic}, &\text{if }
		\alpha_e <\alpha_{\mathrm{y},e}, \\ 
		\text{inelastic}, &\text{otherwise},
	\end{cases} \qquad \text{with } \sigma_\mathrm{com}({\fsig_e}) = \alpha_e;
\end{align}
\item\label{itm:elastic} if $p \equiv \text{elastic}$: \\
assign tangent as elastic stiffness matrix i.e.
\begin{align}
	\fC_e = \fC_e^{\text{el}};
\end{align}

\item\label{itm:inelastic} if $p \equiv \text{inelastic}$:
\begin{enumerate}
	\item set new yield condition
	\begin{align}
		\alpha_{\mathrm{y},e} \equiv \sigma_\mathrm{com} ({\fsig_e});
	\end{align}
	\item find closest data point $(\Delta \hat{\feps}_e, \Delta \hat{\fsig}_e,\hat{\alpha}_e)$  
	by
	\begin{align}
		\argmin\limits_{(\Delta \hat{\feps}_i, \Delta \hat{\fsig}_i, \hat{\alpha}_i)\in \mathcal{D}_e^{\text{ext},\, p}} \|\alpha_e - \hat{\alpha}_i\|_2;
	\end{align}
	\item calculate the diagonal matrix $\fsig^D_e$ of $\fsig_e$ containing principal stresses $\sigma_{1,e} \geq \sigma_{2,e} \geq \sigma_{3,e}$ and the corresponding transformation matrix $\bold T_e$ satisfying $\fsig^D_e = \bold T_e^{-1} \fsig_e \bold T_e$
	\item calculate normal vector in octahedral plane i.e.
	\begin{align}
		\hat{\fN}(\theta_e) = \frac{1}{\sqrt{\rho(\theta_e)^2 + \rho^\prime(\theta_e)^2}} \left[ \fs_e + \sqrt{\frac{2}{3}} \rho^\prime(\theta_e) \begin{pmatrix}
			\sin\left(\theta_e\right) \\ -\cos\left(\frac{\pi}{6}-\theta_e\right) \\ \cos\left(\frac{\pi}{6}+\theta_e\right)
		\end{pmatrix}\right]
	\end{align}
	and transform it into the Cartesian coordinate system 
	\begin{align}
		\fN_e = \bold T_e \cdot \mathrm{diag}(\hat{\fN}) \cdot \bold T_e^{-1};
	\end{align}
	\item determine parameter $\gamma_e$ using the equation
	\begin{align}
		\Delta \hat{\fsig}_e = \lambda_e \mathrm{tr}(\Delta\hat{\feps}_e)\fI + 2\mu_e \Delta\hat{\feps}_e - \gamma_e \fN_e \otimes \fN_e;
	\end{align}
	\item assign tangent as
	\begin{align}
		\fC_e = \fC_e^{\text{el}} + \gamma_e \fN_e \otimes \fN_e;
	\end{align}
\end{enumerate}
\item set the closest data point $(\hat{\fz}_e, \fC_e)$ to modeling point $\fz_e$ as $(\fz_e, \fC_e)$.
\end{enumerate}
\hrule
\vspace{0.5cm}
The first two steps map the transition rules to the octahedral plane and the corresponding data sets. Step \ref{itm:elastic} and \ref{itm:inelastic} assign the tangent stiffness matrix distinguishing between the elastic or inelastic assignment. While Step~\ref{itm:elastic} maps the tangent to the elastic stiffness matrix, Step~ \ref{itm:inelastic} defines a new yield limit and calculates the normal in the octahedral plane, finding the closest point in the data set and using Eq.~\eqref{eq:normal_hw}. The inelastic stiffness matrix \eqref{eq:tangent_eq} is then obtained by a coordinate transformation and Eq.~\eqref{eq:const_incr}. 
\\
Due to the definition of the adapted projection $P_\mathcal{D}$ and the usage of the tangent-space structure in $P_\mathcal{C}$, we conclude that only one fixed-point iteration in Eq.~\eqref{eq:proj_DC} is required. This can be shown assuming a given data state $(\hat{\fz}^{k+1}, \fC^{k+1}) = P_\mathcal{D}(\fz^k)$ with material state $\fz^k = P_\mathcal{C}(\hat{\fz}^k,\fC^k)$ obtained at the $k$-th fixed-point iteration. Using compatibility condition \eqref{eq:constraint1} and equilibrium condition \eqref{eq:constraint2}, it follows that the linear equation system \eqref{eq:lin_eq_sys} of projection $P_\mathcal{C}(\hat{\fz}^{k+1},\fC^{k+1})$ at iteration $k+1$ reads
\begin{align}
\left(\sumE  w_e \fB^T_e \fC_e^{k+1} \fB_e \right) \fu^{k+1} &= \ff - \sumE  w_e \fB^T_e (\hat{\fsig}_e^{k+1}  - \fC_e^{k+1}  \hat{\feps}_e^{k+1} ) \\
&= \ff - \sumE  w_e \fB^T_e (\fsig_e^{k}  - \fC_e^{k+1} \feps_e^{k} )\\
&= \ff - \ff + \sumE  w_e \fB^T_e \fC_e^{k+1} \feps_e^{k} \\
&= \left(\sumE  w_e \fB^T_e \fC_e^{k+1} \fB_e \right) \fu^{k},
\end{align}
which yields $\fu^{k+1} = \fu^{k}$ and therefore $\fz^{k+1} = \fz^{k}$. Consequently, the enhanced paradigm increases efficiency compared to the classical data-driven algorithms  \cite{kirchdoerfer:2016,ciftci:2022}. The detailed adapted data-driven scheme $P_\mathcal{D}P_\mathcal{C}$ is summarized in Algorithm \ref{alg:ddsolver}. The next section demonstrates the performance of the proposed scheme via a numerical example employing elasto-plasticity with isotropic hardening.
\section{Numerical result for a $3D$ benchmark}\label{sec:num_example}
This section illustrates the performance of the adapted data-driven solver extended by the tangential space information in Haigh-Westergaard space in a typical benchmark, considering stress analysis of elasto-plastic material with non-linear isotropic hardening based on experimental measurements synthetically simulated. In our scope, synthetic data consists of strain-stress points created numerically using a material model rather than obtained by actual experimental measurements. We restrict the simulation to noise-free synthetic data sets. However, experimental data is generally noisy and includes outliers. This issue can be treated using noise reduction algorithms such as tensor voting \cite{kim:2013}, Kalman filtering \cite{kalman:1960}, and deep learning-based methods.
\\ \\
The application and the numerical performance will be demonstrated by investigating the impact of increasing the number of tensile data on the convergence of the proposed data-driven scheme. The significance of the quantity of tension-torsion data points is secondary because the accuracy of the approximation of the yield surface is significantly dependent on the method utilized, such as polynomial or spline interpolation, nearest neighbor approaches, or machine learning methods. 
\\
The root-mean-square error between the data-driven strain and stress solution $\fz^k$ and its corresponding reference solution $\fz^{k, \text{ref}}$ will be calculated using
\begin{align}
\text{RMSD}(\fz)^2 &= \frac{\sum_{k=0}^T \text{Error}(\fz^k )^2}{T},
\end{align}
where $T \in \mathbb{N}$ is the number of total loading steps, $\fz^k_e = (\feps_e^k, \fsig_e^k )$ the local data-driven states and  $\fz^{k, \text{ref}}_e = (\feps_e^{k, ref} ,\fsig_e^{k, \text{ref}})$ the local reference states at step $k \leq T$. The error is given by 
\begin{align}
\text{Error}(\fz^k )^2 &=  \frac{\sum_{e=1}^m w_e \|\fz_e^k - \fz^{k, \text{ref}}_e\|_e^2}{ \sum_{e=1}^m w_e \|\fz^{k, \text{ref}}_e\|_e^2},
\end{align}
with $\|\cdot\|_e$ given by the definition in Eq.~\eqref{eq:norm}. The reference states are computed using an iterative return mapping algorithm embedded in a Newton-Raphson global loop restoring equilibrium. 
\subsection{Plate with a circular hole}\label{subsec:plate}
\begin{figure}[h]
\centering
\includegraphics[scale=0.18]{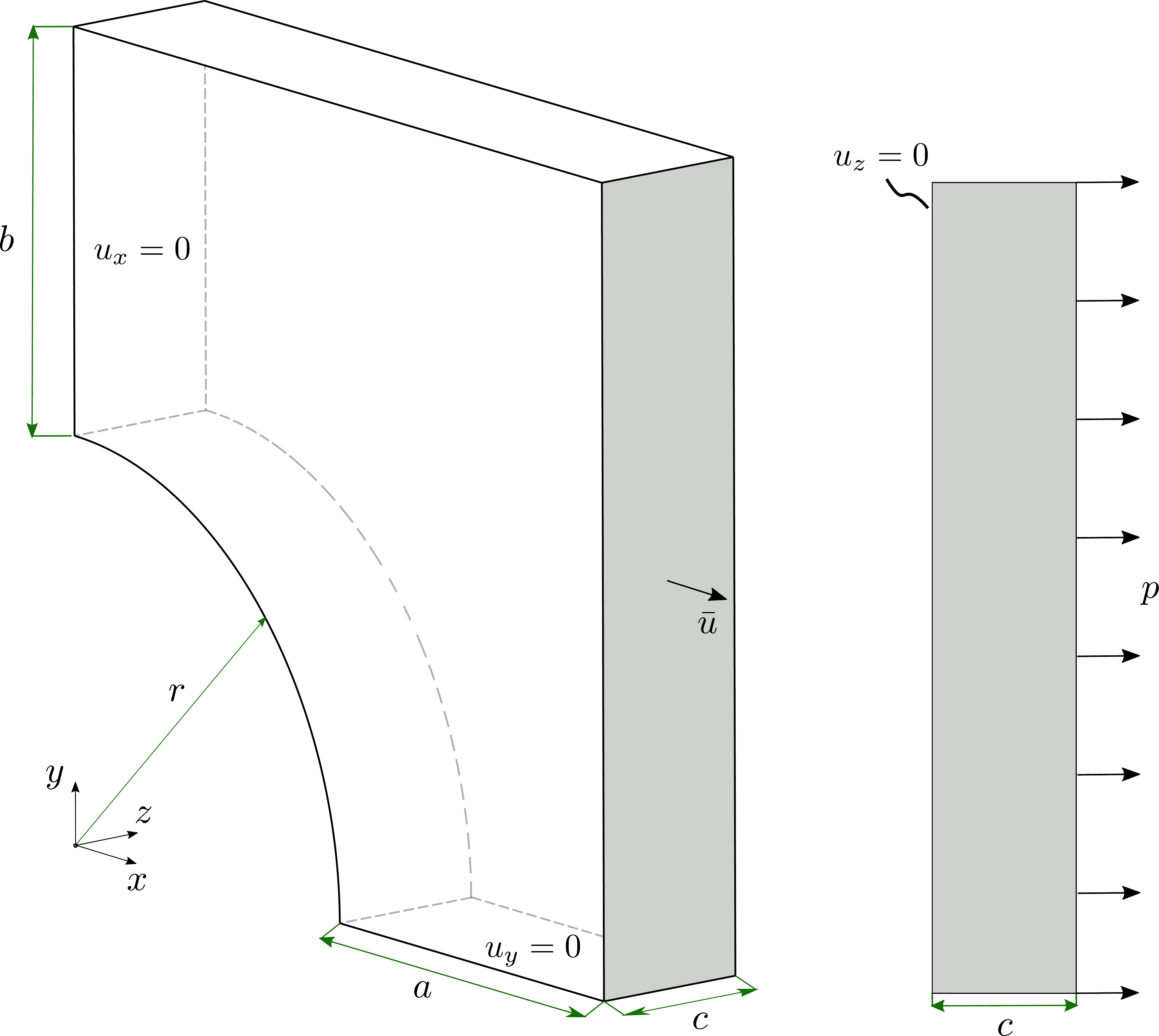}
\caption{Boundary conditions and dimensions of a quadratic plate under increasing extension $\bar{u}$ on the grey area and uniformly distributed pressure $p$ over the thickness direction.}
\label{fig:qplate_geometry}
\end{figure}

In this benchmark, we investigate a $3D$ plate with a circular hole subjected to an increasing extension $\bar{u}$ in the length direction and a uniformly distributed load $p$ over the thickness direction. Due to the symmetry of geometry and load, only one-quarter of the system is modeled. The geometry, boundary conditions, and loading are chosen according to a similar test presented in \cite{peng:2012} and illustrated in Fig. \ref{fig:qplate_geometry}. The side lengths of the strip are equal to $a=b=5\,\text{m}$, and the thickness is $c =2\,\text{m}$. The radius of the hole is $r=5\,\text{m}$. Displacements are fixed at the quarter plates' left surface $x = 0$  in  $x$-direction and at the bottom surface $y = 0$  in $y$-direction. For $z=0$, the displacements are fixed in $z$-direction. The corresponding boundary conditions read as follows:
\begin{equation}
\begin{aligned}\label{eq:qplate_bc}
	\begin{cases}
		u_x = 0,  	& 	\text{if } x = 0;\\
		u_y = 0,  	& 	\text{if } y = 0;\\
		u_z = 0 	& 	\text{if } z = 0;\\
		(u_x, u_y, u_z) = (\bar{u}_x, \bar{u}_y, \bar{u}_z) & 	\text{if } x = a,
	\end{cases}
\end{aligned}
\end{equation}
where $u_x, u_y, u_z$ and $\bar{u}_x, \bar{u}_y, \bar{u}_z$ are the displacements in $x, y$ and $z$-directions, respectively.

\subsubsection*{Material parameters:}\label{subsubsec:plate_matmodel}
\begin{figure}[H]
\centering
\includegraphics[scale=0.8]{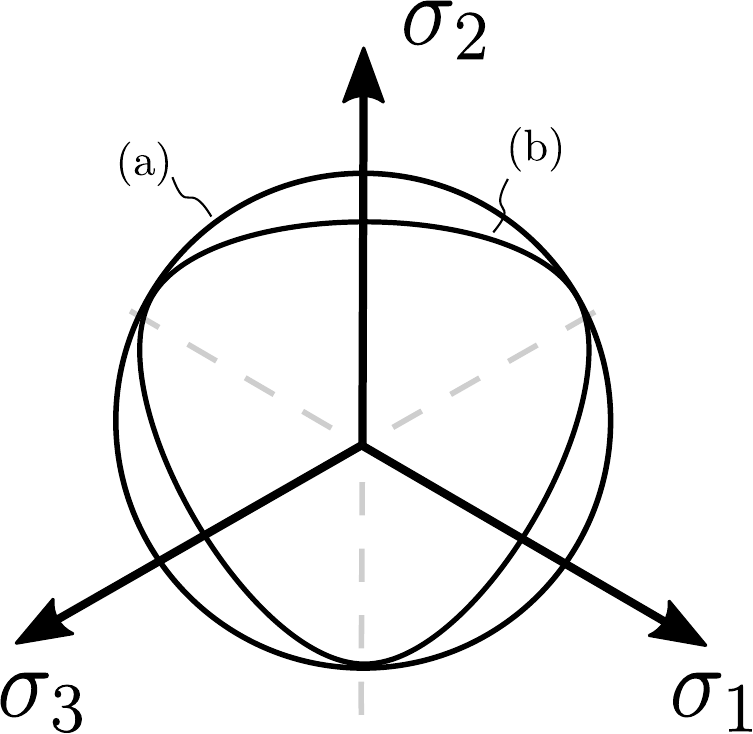}
\caption{Visualization of yield surface function $\Phi(\theta)$ for (a) $k = 1$ and (b) $k = 0.75$.}
\label{fig:yield_num_examples}
\end{figure}

This setting considers an elasto-plastic material with non-linear isotropic hardening. The applied material parameters are Young's modulus $E$, Poisson's ratio $\nu$, and elasticity tensor given by 
\begin{align}\label{eq:exmpl_elas_matrix}
\fC^{\text{el}} = \left(\kappa -  \frac{2}{3}G\right)\fI \otimes \fI + 2G \, \fII,
\end{align}
with bulk and shear moduli $\kappa = \frac{E}{3(1-2\nu)}$ and $G = \frac{E}{2(1+\nu)}$.
A power law describes the non-linear isotropic hardening by
\begin{align}
\sigma_\mathrm{y}(\bar{\varepsilon}^p) =\left(1-\frac{1}{3}\tan 30^\circ\right)\left(\sigma_{\mathrm{0}} + H(\bar{\varepsilon}^p)^{1/h}\right).
\end{align}
where $\sigma_{\mathrm{0}} \in \mathbb{R}_{+}$ is the initial yield limit, $H \in \mathbb{R}_{+}$ is the hardening modulus, $h\in(0,1)$ is the hardening exponent and $\bar{\varepsilon}^p \in  \mathbb{R}_{+}$ defines equivalent plastic strain. For the simulation, we investigate the  yield criterion
\begin{align}\label{eq:yield_kappa}
F(\fsig) \leq \sigma_\mathrm{y}(\bar{\varepsilon}^p),
\end{align}
with
\begin{align}\label{eq:yield_kappa_ineq}
F(\fsig) :=  \frac{\rho\sqrt{3}}{2\sqrt{2}}\left(1+\frac{1}{k}-\left(1-\frac{1}{k}\right)\cos(3\theta)\right),
\end{align}
where $\rho,\theta$  are Haigh-Westergaard Coordinates depending on $\fsig$ and $k \in \mathbb{R}_+$ is a ratio that controls the dependence of the yield surface on the principal stress. One can notice that for $k=1$ Equation~\eqref{eq:yield_kappa} represents a von Mises yield surface, see Fig. \ref{fig:yield_num_examples}, case (a). 
In addition, the initial yield function can be written as
\begin{align}\label{eq:yield_kappa_fct}
\Phi(\theta) = \sigma_\mathrm{y}(0)\left(\frac{\sqrt{3}}{2\sqrt{2}} \left[1+\frac{1}{k}-\left(1-\frac{1}{k}\right)\cos(3\theta)\right]\right)^{-1}.
\end{align}
\subsubsection*{Synthetic data:}\label{subsubsec:plate_data}
The synthetic data in this benchmark setting consists of a combined tension-torsion test, Eq.~\eqref{eq:data_tensile_torsion}, and a single tension test, Eq.~\eqref{eq:data_tensile}. The points $\hat{\fsig}_i$ of the tension-torsion test depend only on the two components $\hat{\sigma}_{11, i}$ and $\hat{\sigma}_{23, i}$ with $i=1,\ldots, n_{1,e}$. Since the data is used to approximate the initial yield surface function $\Phi(\theta)$, it is enough to simulate points lying on it. For this, we define uniformly distributed random components $\hat{\sigma}_{11, i} \in \left[-\sigma_\mathrm{y}(0), k\sigma_\mathrm{y}(0)\right]$. The points $\hat{\fsig}_i$ lying on the yield surface are then determined by $\hat{\sigma}_{11, i}$ and $\hat{\sigma}_{23, i}$ obtained by solving
\begin{align}
\frac{\sqrt{\hat{\sigma}_{11, i}^{2}+3\hat{\sigma}_{23, i}^{2}}}{2k}\left(1+k+\left(1-k\right)\frac{\left(\hat{\sigma}_{11, i}^{3}-9\hat{\sigma}_{11, i}\,\hat{\sigma}_{23, i}^{2}\right)}{\left(\hat{\sigma}_{11, i}^{2}+3\hat{\sigma}_{23, i}^{2}\right)^{\frac{3}{2}}}\right)= \sigma_\mathrm{y}(0).
\end{align}
The second data set is generated by simulating a uniaxial tensile test subject to predefined loading paths. The resulting data points $(\hat{\feps}_i, \hat{\fsig}_i)$ are then used to establish the data set \eqref{eq:dataset_extended_mod} consisting of points $(\Delta\hat{\feps}_i, \Delta \hat{\fsig}_i, \hat{\alpha}_i)$ with $i=1,\ldots, n_{2,e}$. The initial hardening parameter of the data-driven comparison stress in Eq.~\eqref{eq:stress_comp_mod} is given by $\alpha_{y, e} = 1$ for all material states $e=1,\ldots,m$. 
\subsubsection*{Convergence and results:}\label{subsubsec:plate_results}
\begin{figure}[htp]
\centering
\includegraphics[width=.3\textwidth]{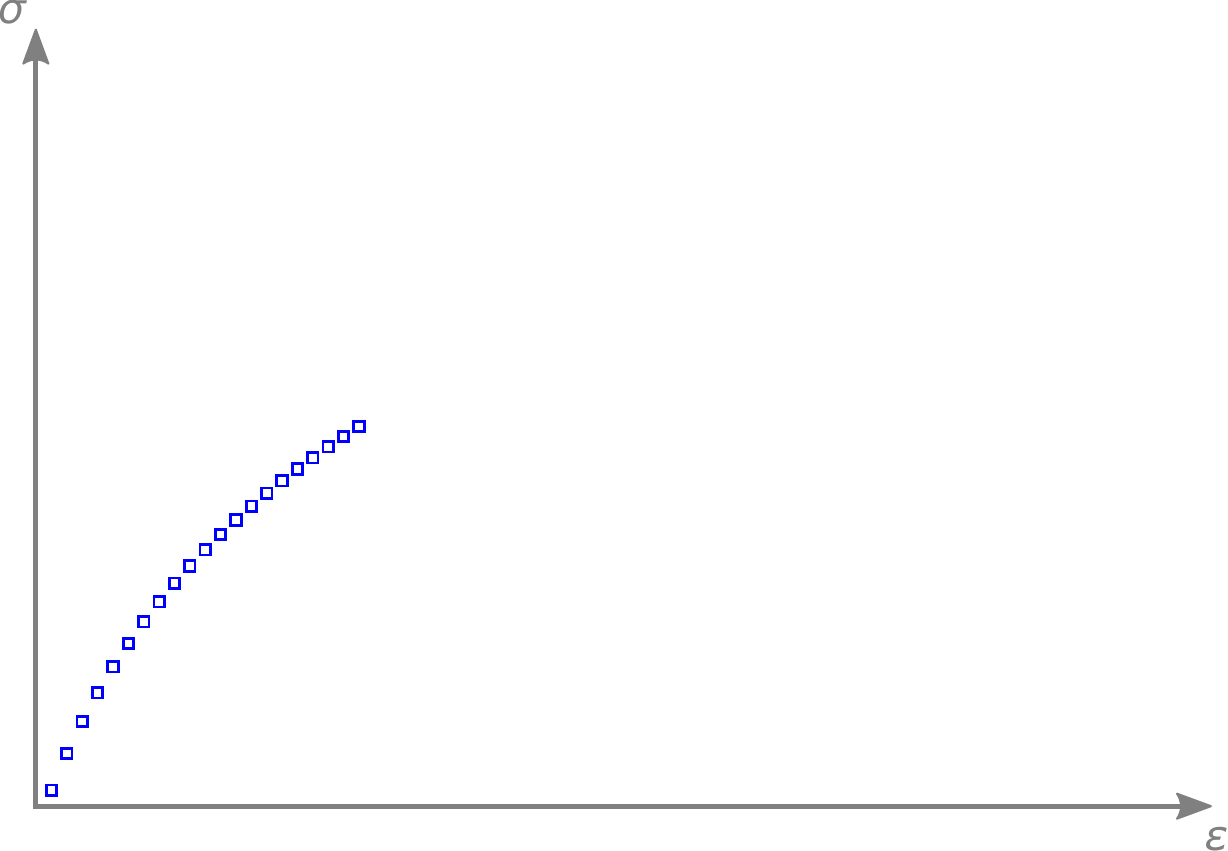}\hfill
\includegraphics[width=.3\textwidth]{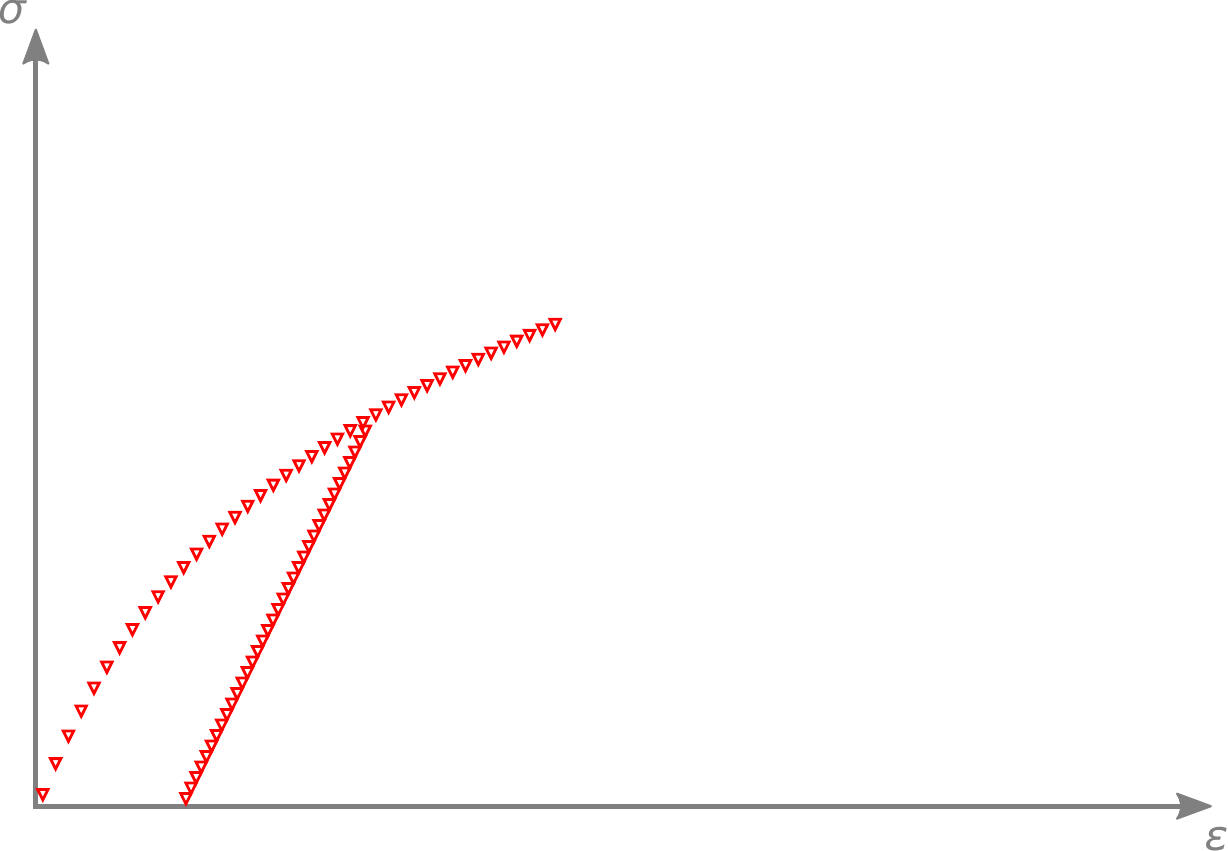}\hfill
\includegraphics[width=.3\textwidth]{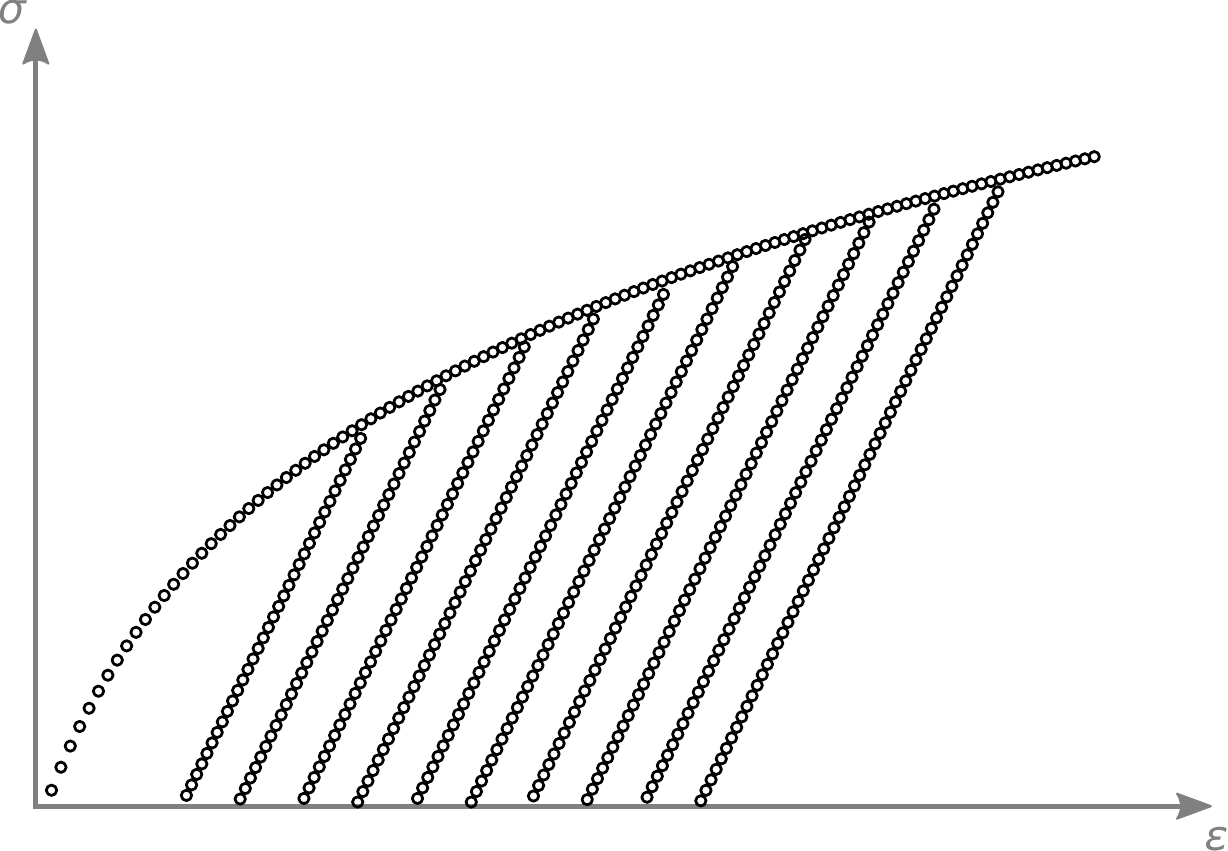}
\caption{An illustrative visualization of synthetic tensile data. The first data set (blue, $\square$) is simulated using a single loading path with 10 data points, and the second data set (red, $\triangle$) consists of two loading paths containing 10 data points. The third data set (black, $\circ$) consists of $10$ loading paths of $10^5$ data points.}
\label{fig:data_sketched}
\end{figure}

The load $p$ increases from $0.0$ to $3.0 \cdot 10^7\, \text{Pa}$, decreases to $0.0 \,\text{Pa}$ and then increases to $3.5 \cdot 10^7\, \text{Pa}$ using $150$ time steps per path with a constant step size of $\Delta t = 1$. Correspondingly, the applied total displacement $\bar{\fu}$ on the lateral side increases to $0.3\,\text{m}$, decreases to $0.0\,\text{m}$ and then increases to $0.4\,\text{m}$. The system is discretized by $10-$nodal tetrahedron elements (P$2$) with quadratic ansatz functions on a mesh of element size $1460$. The exact material parameters used for the reference solution and synthetic data are given in Table \ref{table:plate_mat_parameter}. For $k=0.75$, the yield surface in principle stress plane is depicted in Fig. \ref{fig:yield_num_examples}, case (b). 
\begin{table}[ht]
\centering
\begin{tabular}[t]{lcccccc}
	\hline
	&$E \,[\text{Pa}]$  & $\nu\,[-] $ & $H\,[\text{Pa}]$ &$\sigma_{\mathrm{0}}\,[\text{Pa}]$& $h\,[-]$&$k\,[-]$ \\
	\hline 
	& \\[\dimexpr-\normalbaselineskip+2pt]
	&$3\cdot 10^{10}$& $0.2$ & $2.5\cdot 10^9$ & $3\cdot 10^8$& $2$& $0.75$ \\
	\hline
\end{tabular}
\caption{Material parameters}\label{table:plate_mat_parameter}
\end{table}%
Regarding the data-driven simulation, we fix the number of tensile-torsion data by $n_{1,e} = 50$ and use simple spline interpolation to approximate the yield surface $\Phi(\theta)$. The displacement and maximum principle stress convergence are then investigated for a small tensile data set of size $n_{2,e}=[10,20,40,60,80,90]$ and a more extensive data set of the size of $n_{2,e} = 10^j$ with $j=2, \ldots, 5$. Each set of tensile test data is simulated using a different number of loading paths $n_p$ chosen from the set $ [1,2,4,6,8,10]$. Therefore, we investigate $12 \cdot 6 = 52$ different data sets. For clarity, $n_{2,e} = 10$ and $n_p = 1$ is the data set defined by a single loading path with $10$ data points. The data set consists of two loading paths containing $10$ data points represented by  $n_{2,e} = 10$ and $n_p = 2$. The last data set consists of $10$ loading paths with $10^5$ data points i.e., $n_{2,e} = 10^5$ and $n_p = 10$. Figure.~\ref{fig:data_sketched} illustrates exemplary the generation structure of the synthetic data sets.

\begin{figure}[H]
\centering
\includegraphics[scale=0.68]{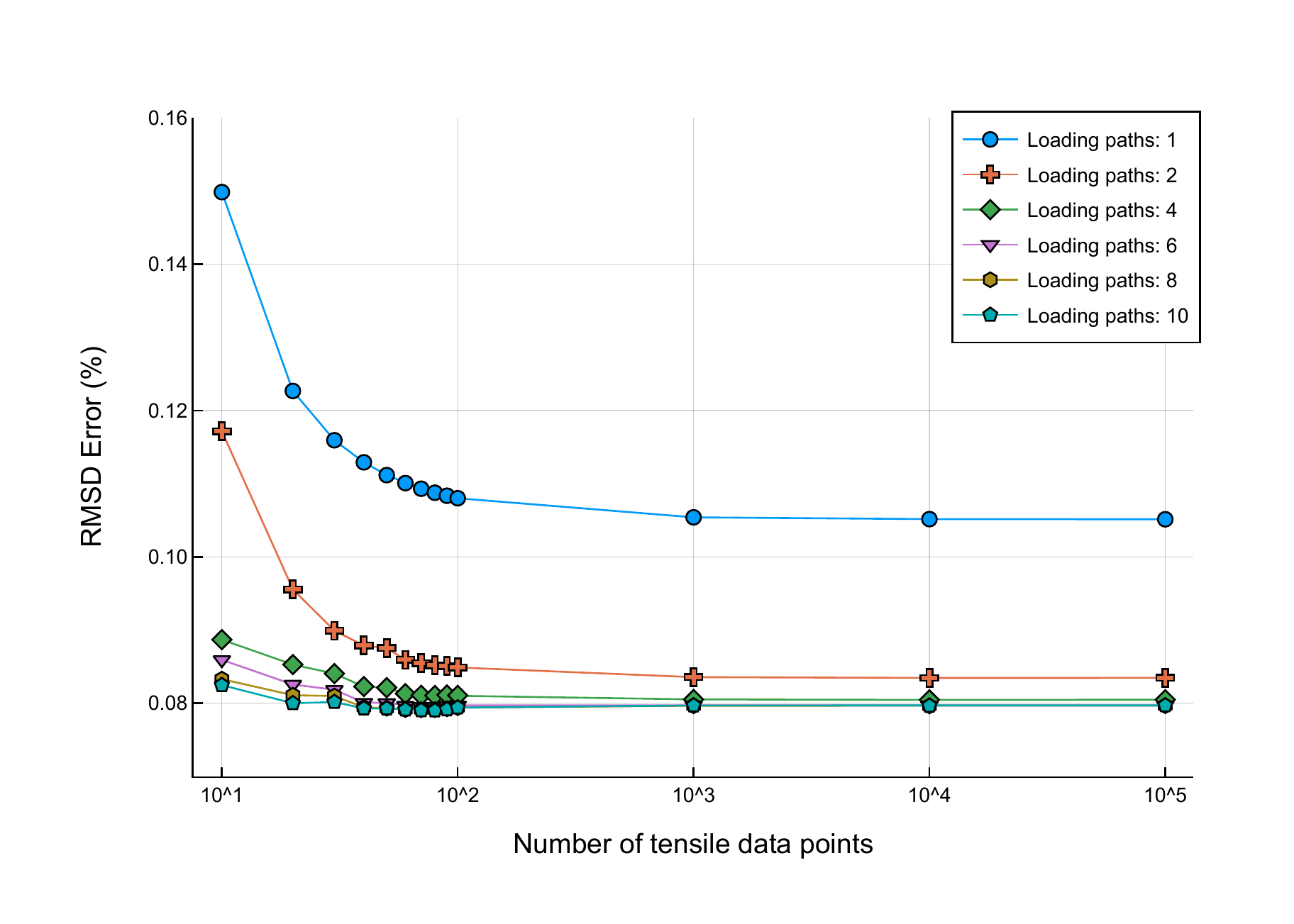}
\caption{RMSD error of the boundary value problem using the adapted data-driven transition mapping to the reference solution based on the exact material model. The graphs are illustrated concerning the size of tensile data and its corresponding number of loading paths.}
\label{fig:rms_error_qplate}
\end{figure}

\begin{figure}[H]
\subfloat[]{{\includegraphics[scale=0.2]{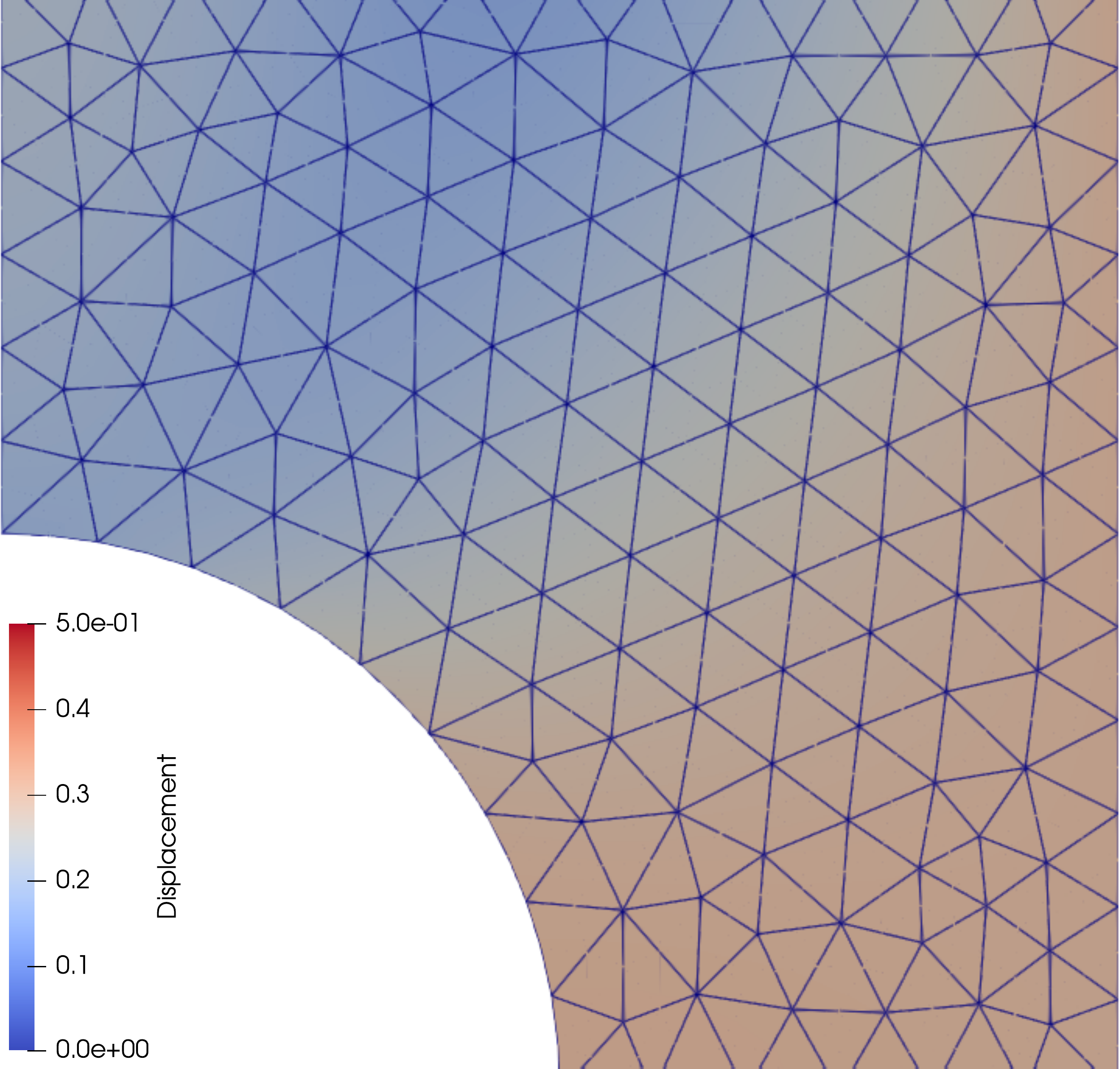}}}
\hspace*{\fill}
\subfloat[]{{\includegraphics[scale=0.2]{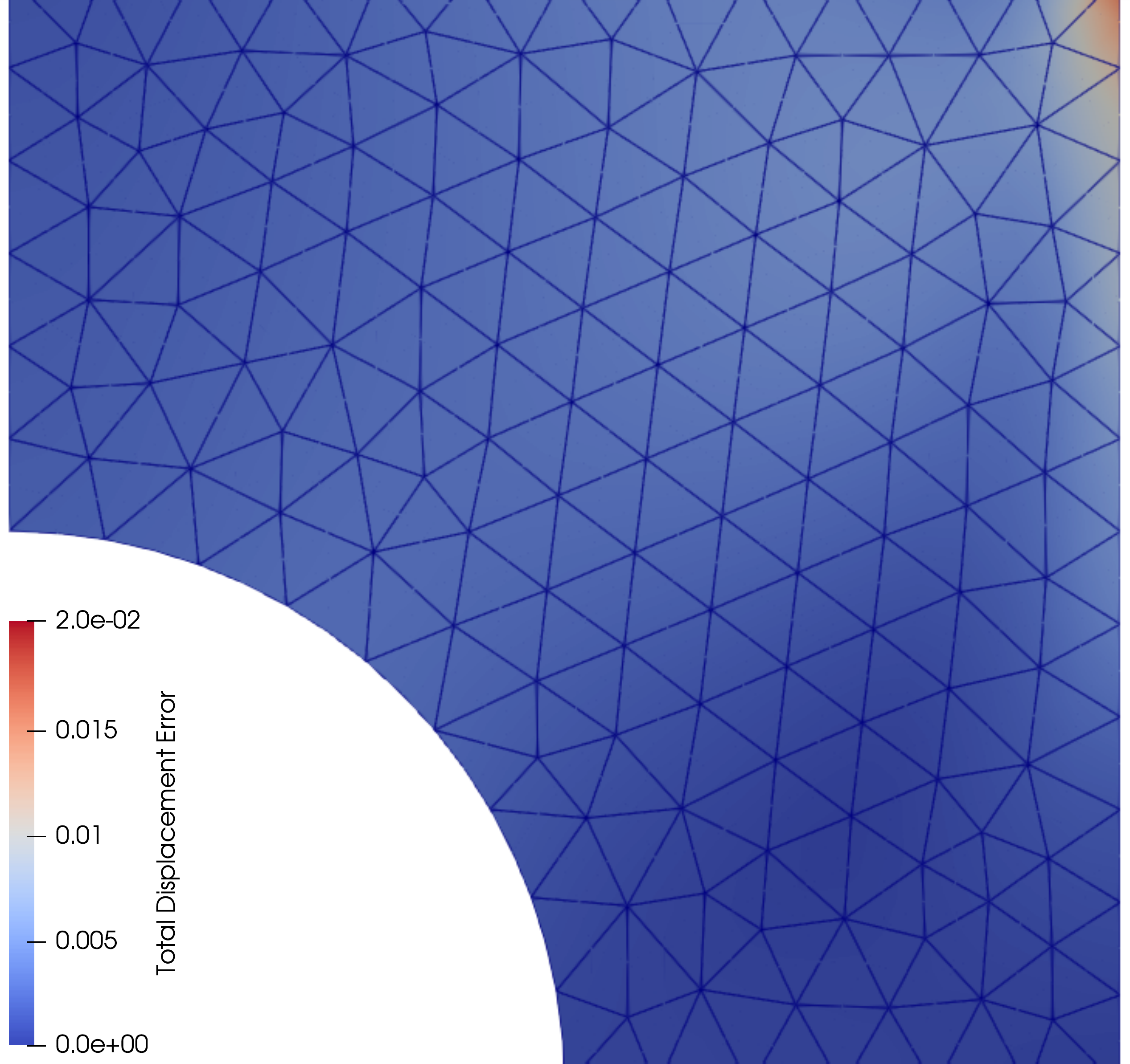}}}%
\\
\subfloat[]{{\includegraphics[scale=0.2]{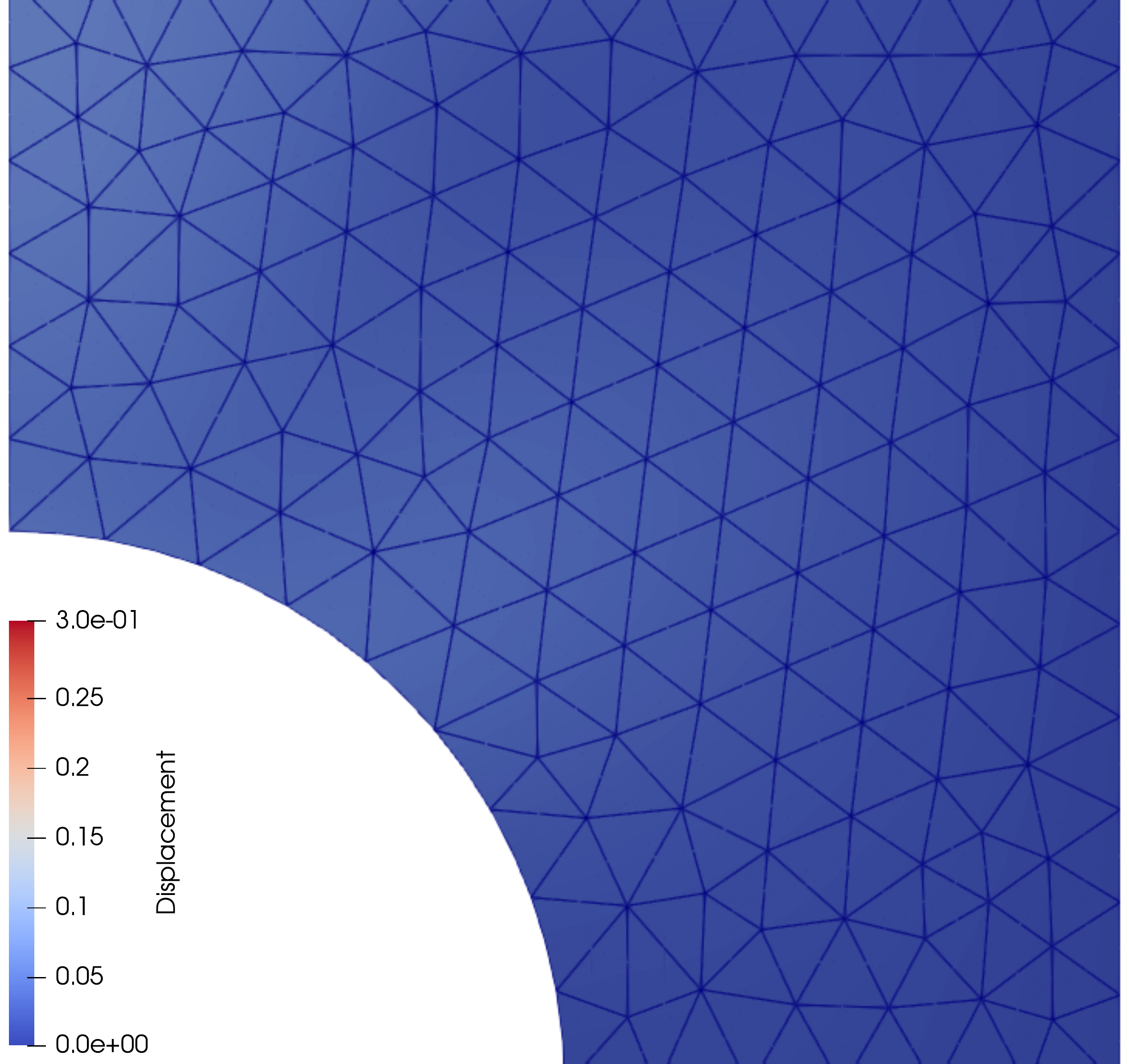}}}
\hspace*{\fill}
\subfloat[]{{\includegraphics[scale=0.2]{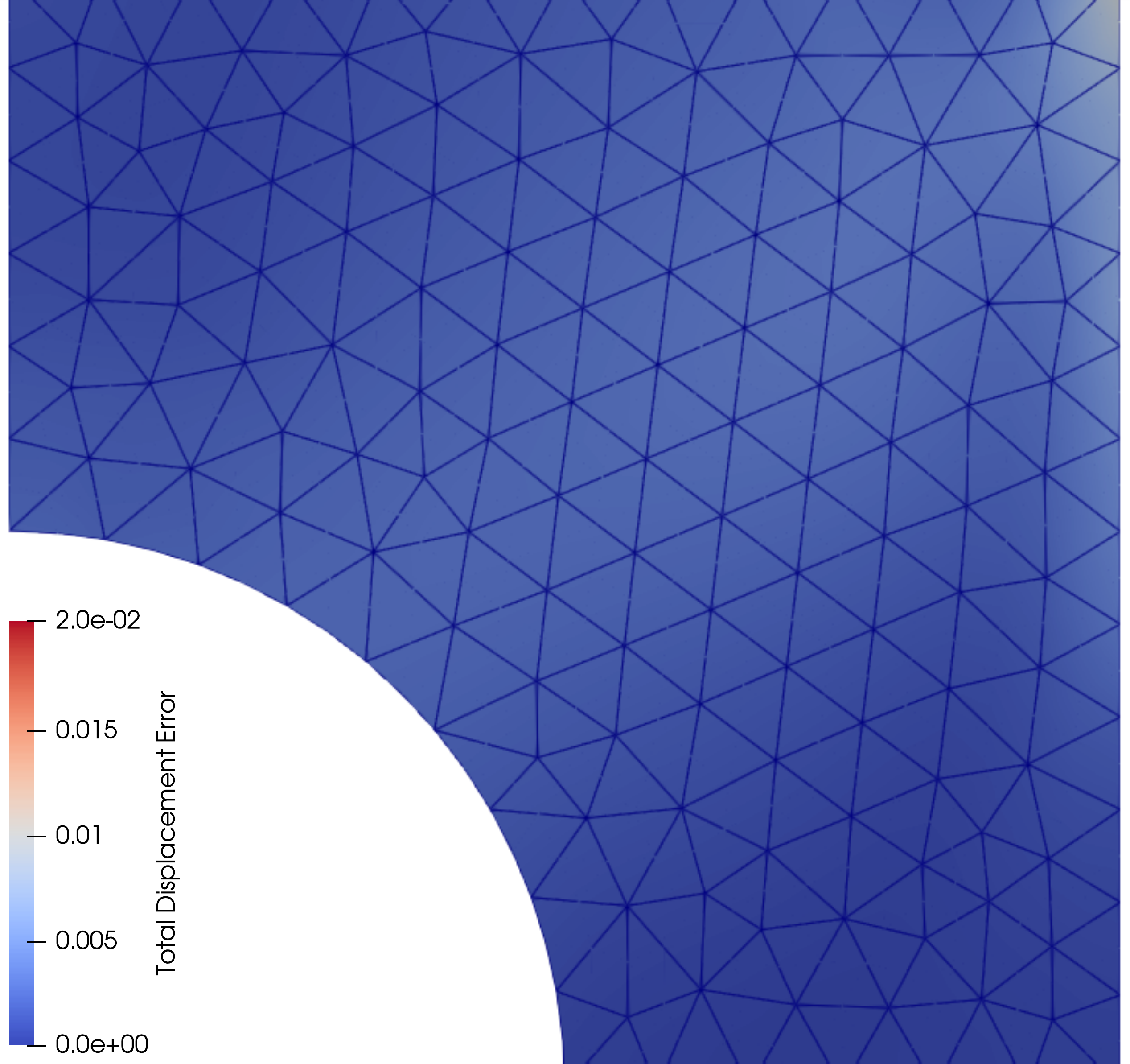}}}%
\\
\subfloat[]{{\includegraphics[scale=0.2]{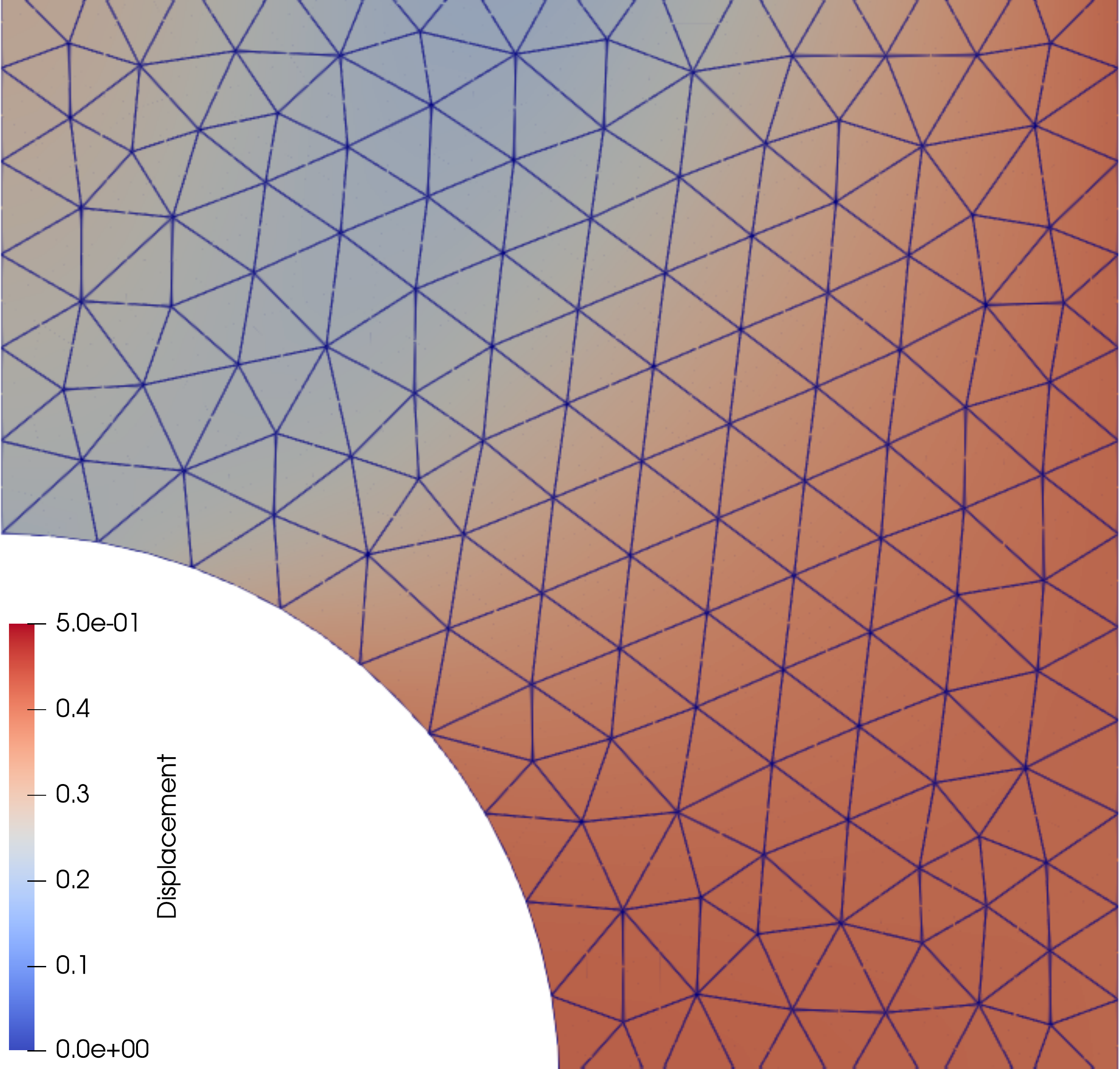}}}
\hspace*{\fill}
\subfloat[]{{\includegraphics[scale=0.2]{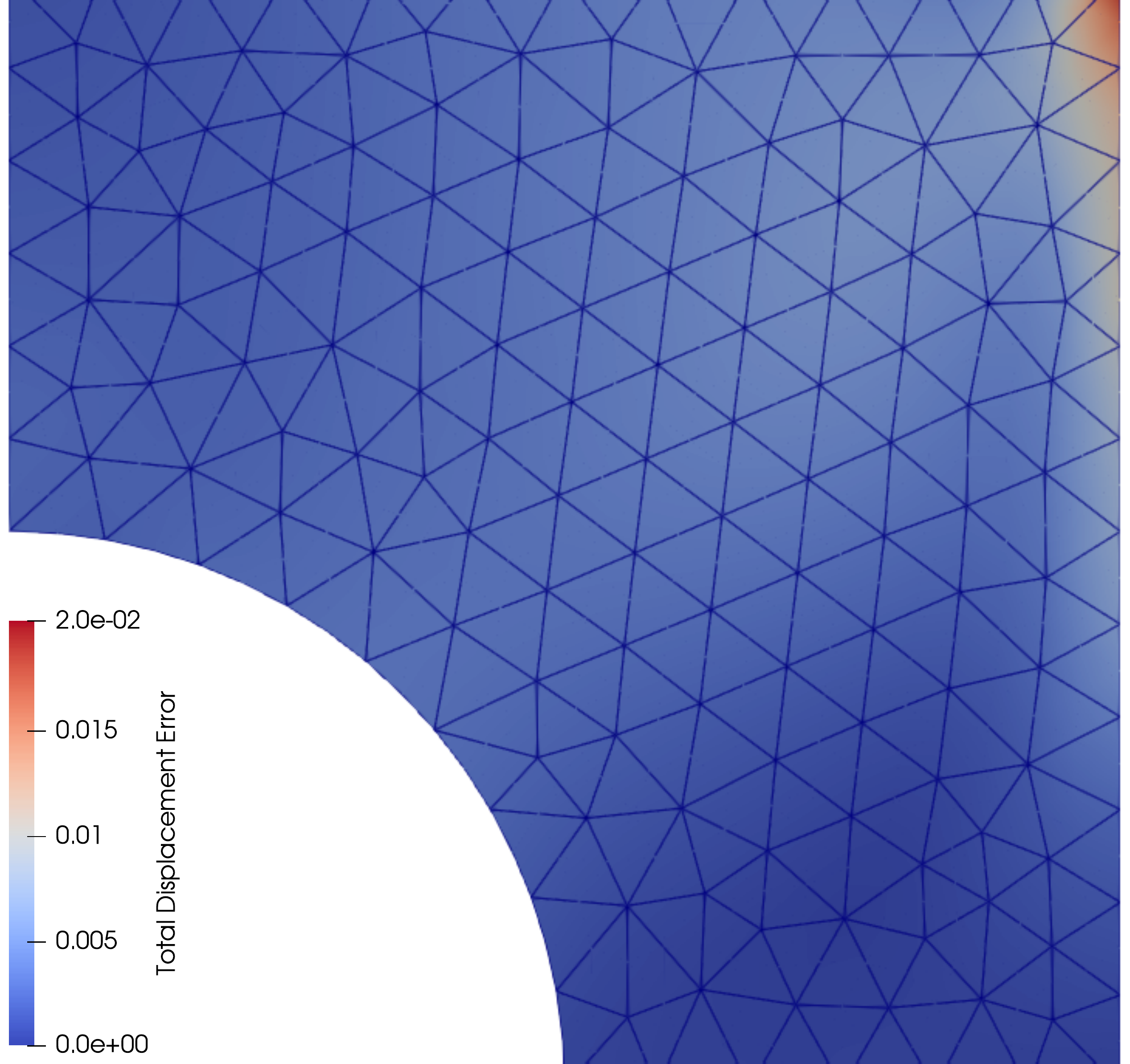}}}
\caption{Contour plot of maximum displacement of the boundary value problem using the adapted data-driven transition mapping. (a, c, e) Maximum displacement and corresponding (b, e, f) absolute errors compared to reference solution at each internal node at time step $t=150, 300, 450$. The number of loading paths simulating the tensile test is $10$ with $10^5$ data points.}
\label{fig:qplate_disp_result}%
\end{figure}

\begin{figure}[H]
\subfloat[]{{\includegraphics[scale=0.2]{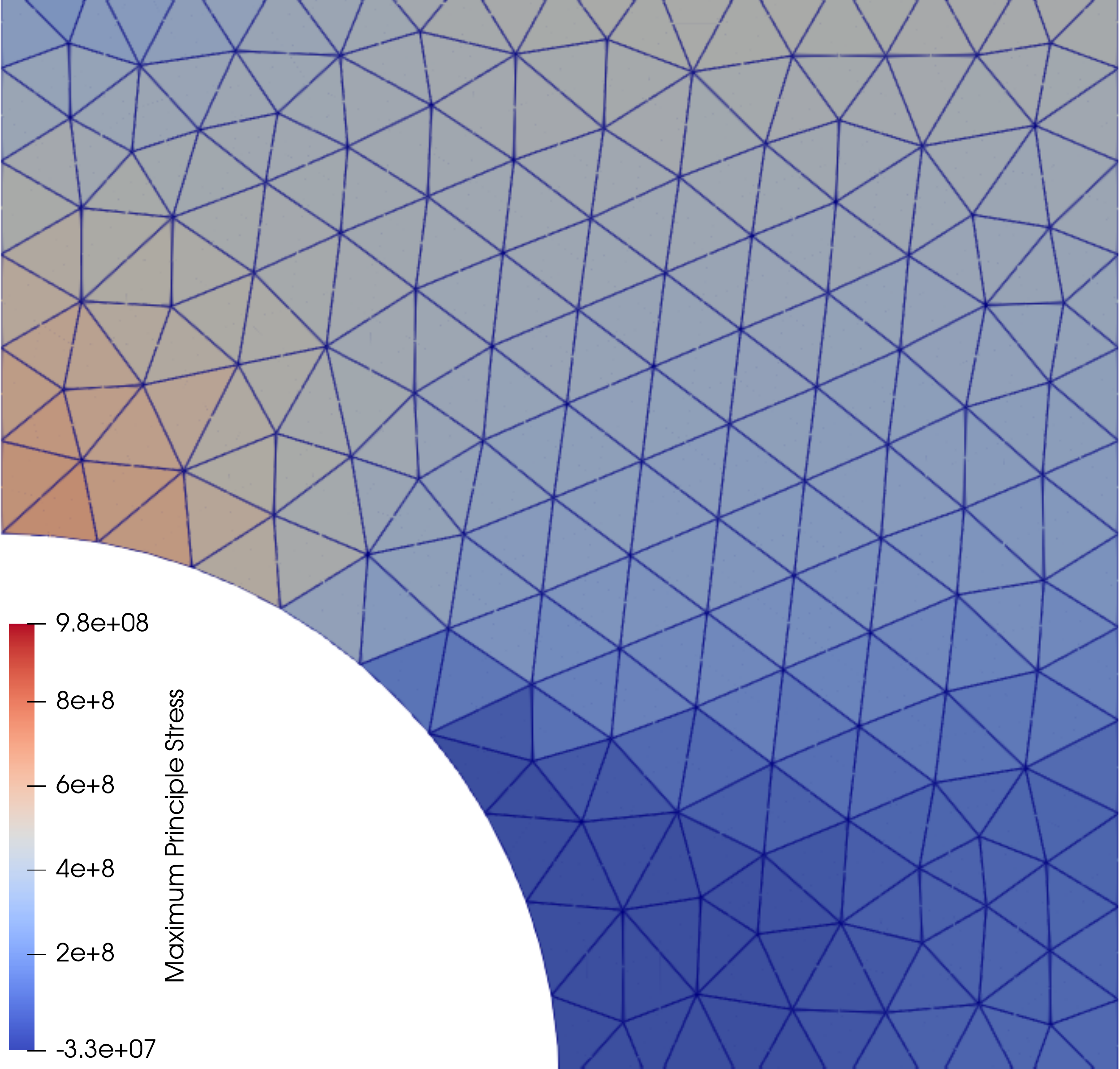}}}
\hspace*{\fill}
\subfloat[]{{\includegraphics[scale=0.2]{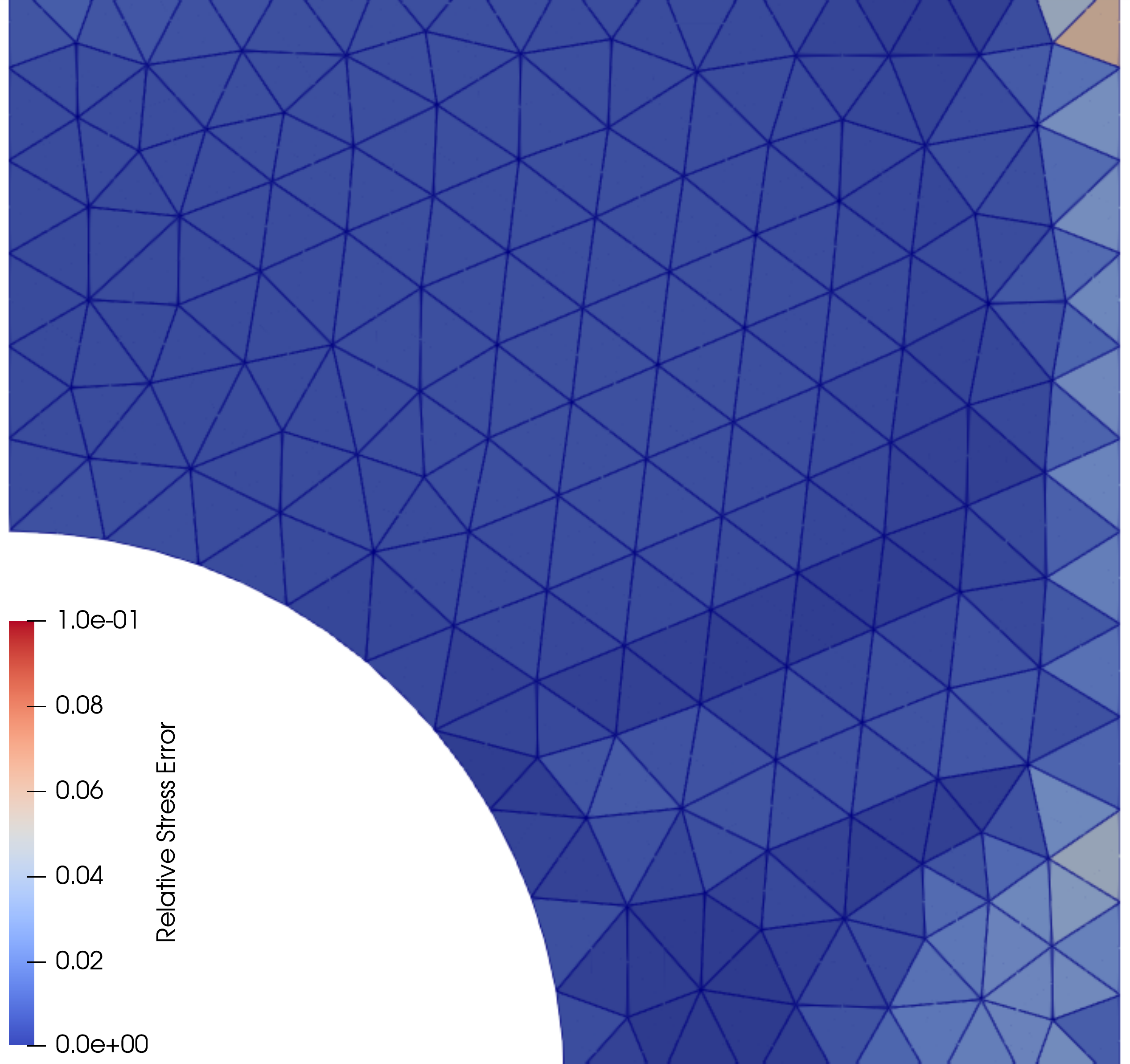}}}%
\\
\subfloat[]{{\includegraphics[scale=0.2]{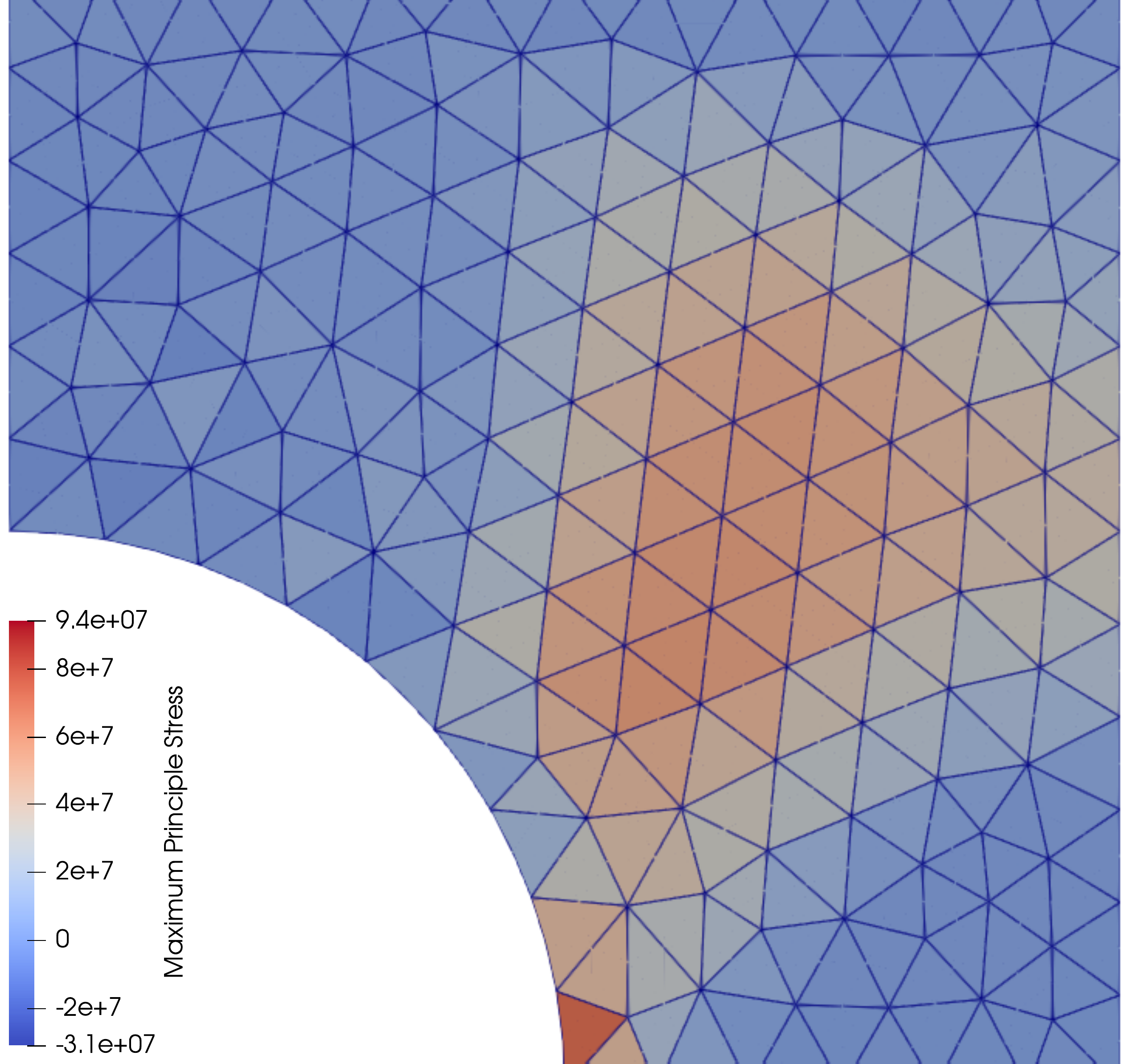} }}
\hspace*{\fill}
\subfloat[]{{\includegraphics[scale=0.2]{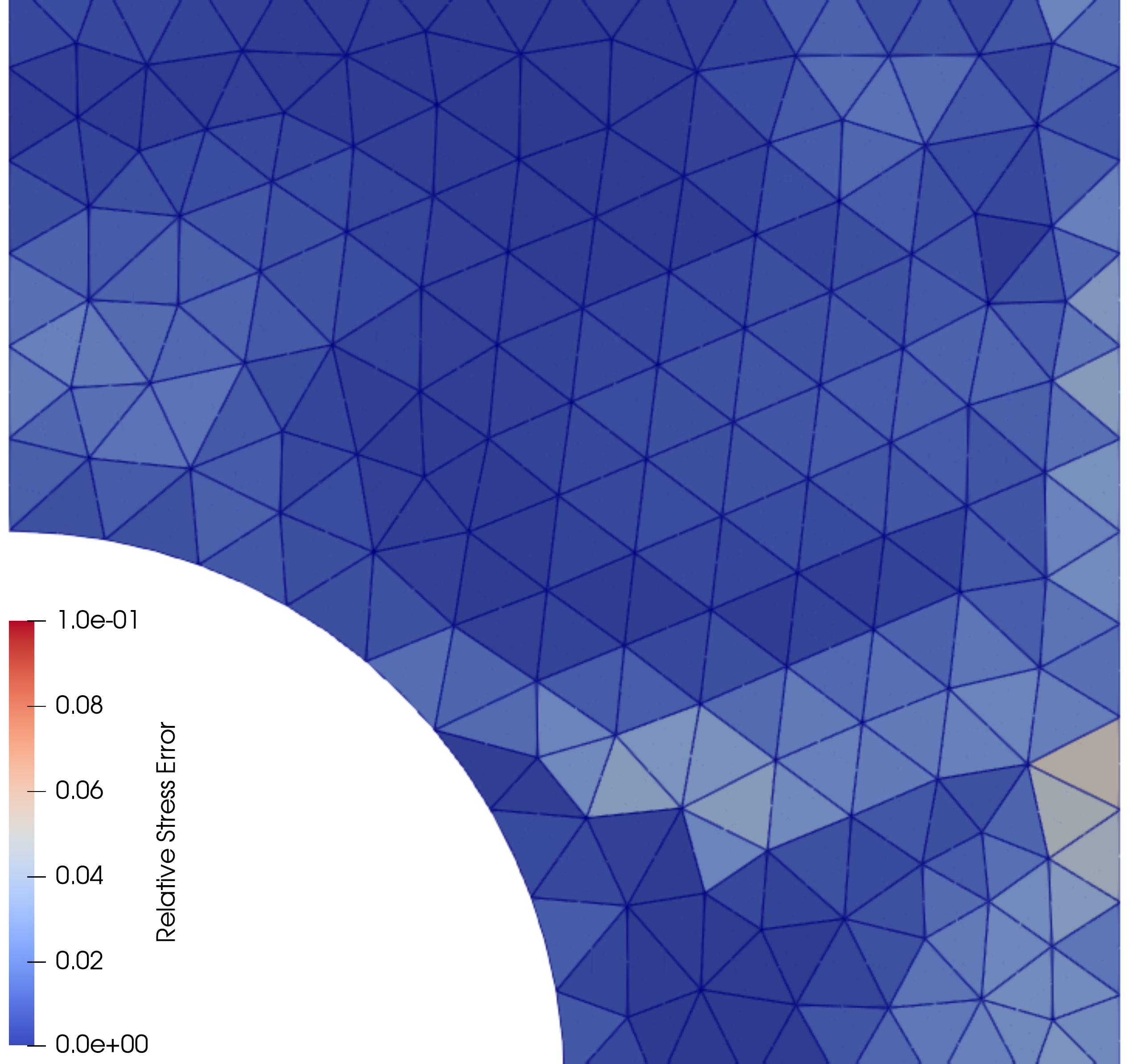} }}
\\
\subfloat[]{{\includegraphics[scale=0.2]{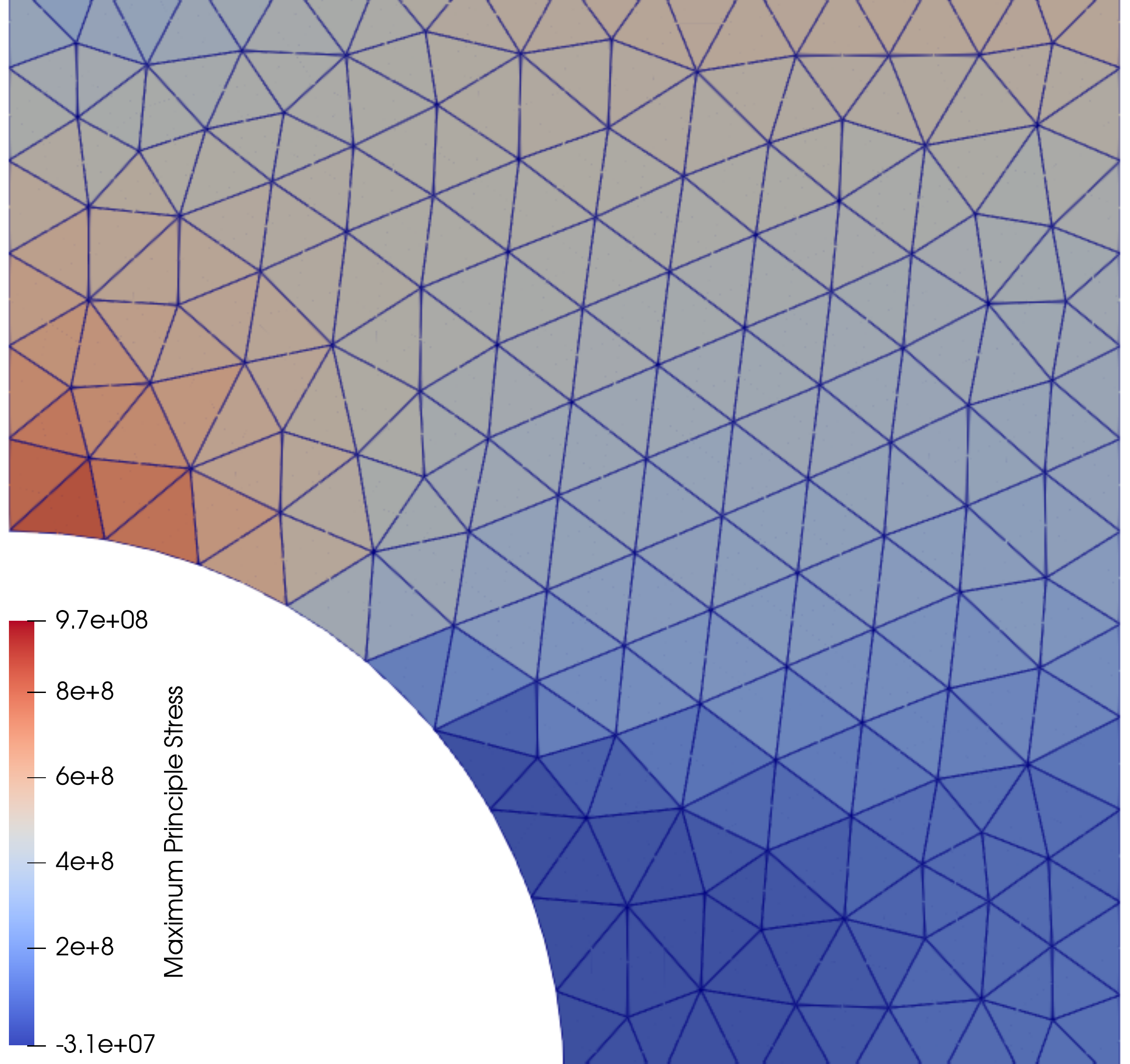} }}
\hspace*{\fill}
\subfloat[]{{\includegraphics[scale=0.2]{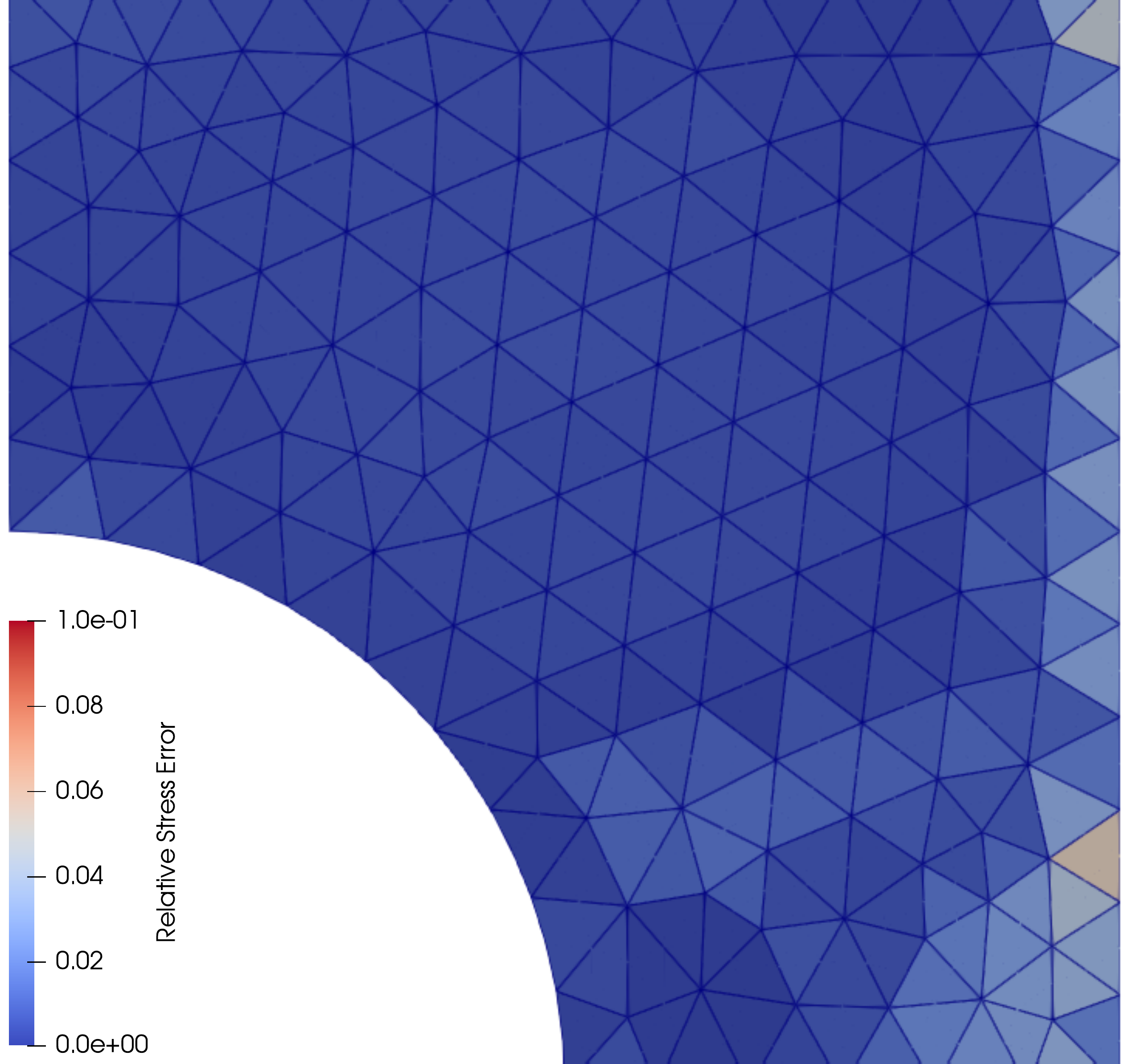} }}

\caption{Contour plot of maximum principal stress of the boundary value problem using the adapted data-driven transition mapping. (a, c, e) Maximum principal stress and corresponding (b, e, f) relative errors compared to reference solution at each material point in [Pa] at time step $t=150, 300, 450$. The number of loading paths simulating the tensile test is $10$ with $10^5$ data points.}
\label{fig:qplate_stress_result}%
\end{figure}

The plot in Fig. \ref{fig:rms_error_qplate} shows the convergence of the error corresponding to the increased data size and the number of loading paths. The bias results from the chosen number of time steps. During the simulation, we noticed that increasing the increment steps decreases the error significantly. Figure \ref{fig:qplate_disp_result} illustrates the occurring displacement in the elasto-plastic plate with a circular hole. In addition, it compares the absolute errors of displacement to the reference solution at internal nodes. Figure \ref{fig:qplate_stress_result} shows the interpolated maximum principle stresses in $[\text{Pa}]$ at each integration point for time step $t=150,300$ and $450$. Furthermore, we plot the relative error compared to the reference solution. We conclude that increasing the number of data and the number of loading paths decreases the RMSD error, corresponding to the data-driven convergence analysis of \cite{ciftci:2022}.
\section{Conclusion}\label{sec:conclusion}
We present using the Haigh-Westergaard space to obtain data points from measurements for the model-free data-driven inelasticity extended by tangent space. Even though the original approach is sufficient for ideal data, the issue of data accessibility and the associated tangent space arises. In particular, data points of inelastic materials could only be acquired through impracticable sample testing in a comprehensive scope of loading directions. 
\smallskip \\
This paper addresses the issue of data accessibility for isotropy guaranteeing the material's loading direction independence. Thus, the tangent space is specified by the material's hardening behavior and the yield surface normal. For the former, we employed data from a straightforward tensile test. We adopted Haigh-Westergaard coordinates for the latter to project the data-driven approach onto the octahedral plane. Then, a combined tension-torsion test provides sufficient information about the underlying material yield surface for approximating the characteristic function. The resulting data-driven method minimizes the distance to the tensile test data and calculates the associated tangent stiffness in the Haigh-Westergaard space, subject to compatibility and equilibrium constraints. The resulting scheme leads to much greater efficiency, especially using material points as data states reduces the fix-point problem to only one iteration. 
\smallskip \\
The application and its numerical performance have been demonstrated on a $3D$ isotropic elasto-plastic benchmark with non-linear hardening. The accuracy improves for larger tensile data sets, and the convergence rate correlates with the convergence analysis of data-driven inelasticity. We neglected the accounting regarding the number of tensile-torsion data points since the quality of the yield surface approximation depends highly on the used method, e.g., polynomial or spline interpolation, nearest neighbor approaches, or machine learning methods. We have limited the simulation to synthetic noise-free data sets. However, experimental data is generally noisy and includes outliers. This issue can be treated using noise reduction algorithms such as tensor voting, Kalman filtering, and deep learning-based methods.
\smallskip  \\
The developments of the data-driven paradigm propose crucial future research areas in machine-learning methods, particularly physics-informed neural networks. By specifying suitable loss functions, these networks can be trained to fulfill training data and discover optimal solutions for given physics-governing equations. Since the data-driven method bypasses the step of material modeling but still relies on solving governing equations, a combined formulation of the model-free data-driven and the physics-informed neural network method is possible.

\appendix
\section{Algorithm of the modified data-driven solver}
\begin{algorithm}[H]
\caption{Data-driven solver at time step $t$  using Haigh-Westergaard coordinates }\label{alg:ddsolver}
\begin{algorithmic} 
	\Require matrices $\{\fB_e\}_{e=1}^m$, weights $\{w_e\}_{e=1}^m$, load $\ff$, Lamé constants $\lambda_e, \mu_e$
	\Data tensile data $\mathcal{D}_e^\text{ext}$, $\Phi_e(\theta)$ obtained through tension-torsion data $\{(\hat{\rho}_i, \hat{\theta}_i)\}_{i=1}^{n_e}$
	\Procedure{Data-Driven Solver}{}
	\If{$t=1$} \Comment{\textbf{Initialize variables}}
	\ForAll{$e=1,\ldots,m$}
	\State $\fC_e^\mathrm{el} = \lambda_e \fI \otimes \fI + 2\mu_e \fII$ \Comment{Elastic stiffness}
	\State $\hat{\fz}_e = ((\hat{\feps}_e,\hat{\fsig}_e), \fC_e) \leftarrow ((\bold 0, \bold 0), \fC_e^\mathrm{el})$  \Comment{Data state}
	\State $\alpha_{y, e} \leftarrow 1$ \Comment{Transition variable}
	\EndFor
	\EndIf
	\State \Comment{\textbf{Projection} $P_\mathcal{D}P_\mathcal{C}$}
	\State $\{\fz_e\}_{e=1}^{m} = \Call{$P_\mathcal{C}$}{\{\hat{\fz}_e\}_{e=1}^{m}}$ 
	\State $\{\hat{\fz}_e, \fC_e\}_{e=1}^m$ = \Call{$P_\mathcal{D}$}{$\{\fz_e\}_{e=1}^{m}$}
	\State $t \leftarrow t+1$
	\EndProcedure%
\end{algorithmic}
\end{algorithm}

\begin{algorithm}[H]
\caption*{\textbf{Projection} $P_\mathcal{C}(\hat{\fz})$}
\begin{algorithmic}     
	\State Solve equation system: 
	\begin{align*}
		\left(\sumE  w_e \fB^T_e \fC_e \fB_e \right) \fu = \ff - \sumE  w_e \fB^T_e (\hat{\fsig}_e - \fC_e \hat{\feps}_e)
	\end{align*}
	\ForAll{$e = 1, \ldots, m$}
	\State $\feps_e = \fB_e \fu,$
	\State $\fsig_e = \hat{\fsig}_e + \fC_e(\feps_e - \hat{\feps}_e)$
	\EndFor
	\State \Return $\{\fz_e\}_{e=1}^{m}=\{(\feps_e, \fsig_e)\}_{e=1}^{m}$
\end{algorithmic}
\end{algorithm}

\begin{algorithm}[H]
\caption*{\textbf{Projection} $P_\mathcal{D}(\fz)$}
\begin{algorithmic}     
	\ForAll{$e=1, \ldots, m$}
	\State $\fsig_e^D, \bold T_e = \Call{PrincipalStress}{\fsig_e}$ \Comment{Diagonalization}
	\State $(\rho_e, \theta_e)  \leftarrow \left(2\sqrt{J_{2, e}}, \frac{1}{3}\arccos{\left(\frac{3\sqrt{2}}{2} J_{3, e} J_{2, e}^{-3/2}\right)}\right)$ 	\Comment{Haigh–Westergaard}
	\State $\alpha_e = \frac{\rho_e}{\Phi_e(\theta_e)}$
	\vspace{0.2cm}
	\If{$\alpha_e \leq \alpha_{\mathrm{y}, e}$} \Comment{Transition rule}
	\State $\fC_e \leftarrow \fC_e^\text{el}$
	\Else
	\State $(\Delta \hat{\feps}_e, \Delta \hat{\fsig}_e, \hat{\alpha}_e) = \argmin\limits_{(\Delta \hat{\feps}_i, \Delta \hat{\fsig}_i, \hat{\alpha}_i)\in\hat{\mathcal{D}}_e^\text{ext}} \|\alpha_e - \hat{\alpha}_i\|_2$
	\State $\rho^\prime_e = \alpha_e \Phi_e^\prime(\theta_e)$
	\State $\fN_e, \gamma_e = \Call{Normal}{(\rho_e, \rho^\prime_e, \theta_e), (\Delta \hat{\feps}_e, \Delta \hat{\fsig}_e, \hat{\alpha}_e), \bold T_e, \lambda_e, \mu_e}$
	\State $\fC_e \leftarrow \fC_e^\text{el} - \gamma_e\fN_e\otimes\fN_e$
	\State $\alpha_{\mathrm{y}, e} \leftarrow \alpha_e$
	\EndIf
	\EndFor
	\State \Return 	$\{(\fz_e, \fC_e)\}_{e=1}^{m}$
\end{algorithmic}
\end{algorithm}

\begin{algorithm}[H]
\caption*{\textbf{Functions} }
\begin{algorithmic}
	\Function{PrincipalStress}{$\fsig$}
	\State \Return diagonal matrix $\fsig^D$ containing principal stresses $\sigma_1 \geq \sigma_2 \geq \sigma_3$ and corresponding transformation matrix $\bold T$ satisfying $\fsig^D = \bold T^{-1} \fsig \bold T$
	\EndFunction
	\Function{Normal}{$(\rho, \rho^\prime, \theta), (\Delta \hat{\feps}, \Delta \hat{\fsig}, \hat{\alpha}), \bold T, \lambda, \mu$}
	\State $\hat{\fN} =\frac{\sqrt{2}}{\sqrt{3}\sqrt{\rho^2 + \rho^{\prime 2}}} \left[\rho \begin{pmatrix}
		\cos\left(\theta\right) \\ 
		\cos\left(\theta - \frac{2\pi}{3}\right) \\
		\cos\left(\theta + \frac{2\pi}{3}\right)
	\end{pmatrix} + 
	\rho^\prime \begin{pmatrix}
		\sin\left(\theta\right) \\ -\cos\left(\frac{\pi}{6}-\theta\right) \\ \cos\left(\frac{\pi}{6}+\theta\right)
	\end{pmatrix}\right]$
	\State $\fN \leftarrow \bold T \cdot \mathrm{diag}(\hat{\fN}) \cdot \bold T^{-1}$
	\vspace{0.2cm}
	\State $\gamma \leftarrow \Delta \hat{\fsig} = \lambda \mathrm{tr}(\Delta\hat{\feps})\fI + 2\mu \Delta\hat{\feps} -\gamma \fN \otimes \fN,$
	\vspace{0.2cm}
	\State \Return $\fN, \gamma$
	\EndFunction
\end{algorithmic}
\end{algorithm}

\end{document}